\documentclass[preprint2]{aastex}

\usepackage{graphicx}

\shorttitle{A unified solution for the orbit and light-time effect in the V505 Sgr system}
\shortauthors{M.~Bro\v z et al.}

\begin{document}

\title{A unified solution for the orbit and light-time effect in the V505 Sgr system}

\author{M.~Bro\v z and P.~Mayer}
\affil{Institute of Astronomy, Charles University, Prague, V Hole\v sovi\v ck\'ach 2, 18000 Prague 8, Czech Republic}

\author{T.~Pribulla\altaffilmark{1}}
\affil{Astrophysikalisches Institut und Universit\"ats-Sternwarte, Schillerg\"asschen 2-3, D-07740 Jena, Germany}

\author{P.~Zasche and D. Vokrouhlick\'y}
\affil{Institute of Astronomy, Charles University, Prague, V Hole\v sovi\v ck\'ach 2, 18000 Prague 8, Czech Republic}

\and

\author{R.~Uhl\'a\v r}
\affil{Private observatory, Poho\v r\'i 71, 25401 J\'ilov\'e u Prahy, Czech Republic}

\altaffiltext{1}{On leave from Astronomical Institute, Slovak Academy of Sciences, 05960 Tatransk\'a Lomnica, Slovakia}



\begin{abstract}
The multiple system V505~Sagittarii is composed of at least three stars:
a compact eclipsing pair and a distant component, which orbit is measured
directly using speckle interferometry.
In order to explain the observed orbit of the third body in V505~Sagittarii
and also other observable quantities, namely the minima timings of the eclipsing binary
and two different radial velocities in the spectrum,
we thoroughly test a fourth-body hypothesis --- a perturbation by a dim, yet-unobserved object.
We use an N-body numerical integrator to simulate future and past orbital evolution
of 3 or 4 components in this system. We construct a suitable $\chi^2$ metric from all available
speckle-interferometry, minima-timings and radial-velocity data and we scan
a part of a parameter space to get at least some of allowed solutions.
In principle, we are able to explain all observable quantities by a presence of a fourth body,
but the resulting likelihood of this hypothesis is very low.
We also discuss other theoretical explanations of the minima timings variations.
Further observations of the minima timings during the next decade
or high-resolution spectroscopic data can significantly constrain the model.
\end{abstract}



\section{Introduction}

The star V505 Sagittarii (HD~187949, HR~7571, HIP~97849, WDS~19531-1436)
is known as an eclipsing binary with a variable period.
Spectral types of its primary and secondary components are A2\,V and G5\,IV,
orbital period is 1\fd183 and visual magnitude in maximum 6\fm5 (Chambliss et al. 1993).
In 1985, the V505 Sgr was also resolved using speckle interferometry (McAlister et al. 1987a)
and several measurements of this 3rd component were published since that time.
Mayer (1997) attempted to join the measured times of minima with
visual orbit and determined a distance of the system 102\,pc.

The 3rd-body orbit with the period of about 40 years seemed well justified
until about the year 2000. An abrupt change in more recent data however
excludes this simple model --- it is impossible to fit {\em both\/} light-time effect data 
and the interferometric trajectory assuming three bodies on stable orbits.
We thus test a 4th-body hypothesis: a perturbation by low-mass star
(i.e., the 4th body), which is not yet visible in the speckle-interferometry data.
Such a fourth body was suspected already by Chochol et al.~(2006)
due to conspicuous deviations of minimum times from expectations.
While we consider the 4th-body model as the main working hypothesis
in this paper, we also discuss other possible effects that can produce
minima timings variations.

The dataset we have for V505~Sgr is described in Section~\ref{sec:data}.
We introduce our dynamical model, numerical method, free/dependent parameters
and $\chi^2$ metric in Section~\ref{sec:integrator}.
The results of our simulations and conclusions are presented in
Sections~\ref{sec:results} and~\ref{sec:conclusions}.


\section{Observational data}\label{sec:data}

\subsection{Speckle interferometry}

The available speckle-interferometry data are summarised in Table~\ref{tab:sky}.
Most of them were extracted from the Fourth Catalog of Interferometric
Measurements of Binary Stars (Hartkopf et al. 2009), but we also
added two speckle measurements from SAO BTA 6\,m telescope by E.~Malogolovets
(using a speckle camera and a method described in Balega et al. (2002) and Maksimov et al. (2009))
and one direct-imaging measurement, performed at CFHT by S.~Rucinski
(using a method described in Rucinski et al. (2007)).

We estimated weight factors $w$ and corresponding uncertainties as
$\sigma_{\rm sky} = 0.005\,{\rm arcsec}/w$. The values of $\sigma_{\rm sky}$
vary because of different telescopes and techniques were used.
Note these measurement errors sometimes cause that some measured points
are seemingly `exchanged' --- the position angle does not revolve monotonically.

We are aware of a possible $180^\circ$-ambiguity in the speckle measurements,
but V505 Sgr is a lucky case: we have one direct measurement by Hipparcos prior
to 2000 perihelion passage, and another direct-imaging datum after 2000.
We thus can be sure about the shape of the orbit.


\begin{table*}
\caption{Speckle interferometry data for V505~Sgr, mainly from the 4th Interferometric Catalogue (Hartkopf et al. 2009).
P.A., $d$ denote position angle and angular distance between the central pair (1+2) and the 3rd component.
Estimated weight factors $w$ and uncertainties $\sigma_{\rm sky} = 0.005\,{\rm arcsec}/w$
correspond to the sizes of telescopes and techniques, which were used to acquire
these measurements (1991.25 and 2005 measurements result from direct imaging).}
\label{tab:sky}
\centering
\begin{tabular}{cllll}
\tableline\tableline
  year & P.A. / deg & $d$ / mas & weight &  source \\
\tableline
  1985.5150  &  189.6  &   302   &   1    &  3.6 m \cr
  1985.8425  &  189.8  &   311   &   1    &  3.8 m \cr
  1989.3069  &  181.0  &   261   &   1    &  4.0 m \cr
  1990.3445  &  176.9  &   246   &   1    &  4.0 m \cr
  1991.2500  &  170    &   234   &   0.6  &  Hipparcos \cr
  1991.3903  &  173.4  &   234   &   1    &  4.0 m \cr
  1991.5575  &  174    &   240   &   0.4  &  2.1 m \cr
  1991.5602  &  174    &   260   &   0.4  &  2.1 m \cr
  1991.7124  &  173.3  &   226   &   1    &  4.0 m \cr 
  1992.4497  &  171.7  &   214   &   1    &  4.0 m \cr
  1992.6961  &  164    &   190   &   0.4  &  2.1 m \cr
  1994.7079  &  159.9  &   192   &   1    &  3.8 m \cr
  1995.4398  &  152.5  &   169   &   0.6  &  2.5 m \cr
  1995.7675  &  154.2  &   177   &   0.3  &  2.5 m \cr
  1996.5320  &  145.8  &   149   &   0.3  &  2.5 m \cr

  2003.6365  &  236.3  &   152   &   1    &  3.5 m\cr
  2005.7948  &  218    &   183   &   0.6  &  direct CFHT \cr
  2006.1947  &  215.8  &   182   &   1    &  4.0 m\cr
  2007.3306  &  212.4  &   210   &   1    &  3.5 m\cr
  2007.4927  &  212.0  &   212   &   1    &  6.0 m\cr
  2008.4901  &  207.8  &   231   &   1    &  6.0 m\cr
  2009.2662  &  204.2  &   247.5 &   1    &  4.0 m\cr
\tableline
\end{tabular}
\end{table*}


\subsection{Minima timings}\label{sec:lite_data}

We list recent $O-C$ data for the (1+2) binary in Table~\ref{tab:lite}.
Only measurements not presented in Chambliss et al. (1993) are included in the table,
but we use all of them of course.
An uncertainty of a minimum determination is estimated to $\sigma_{\rm lite} = 1\,{\rm min}$
in most cases, only photographic minima and data from Hipparcos were considered worse.
Epoch and $O-C$ were calculated using the ephemeris:
\begin{equation}
{\rm Pri.Min.} = 2433490.483 + 1\fd1828688 \times E\label{eq:ephem}
\end{equation}
Note there is a freedom in period and base minimum determination.
When we compare these $O-C$ measurements to our simulations we use
an optimal ephemeris, different from~(\ref{eq:ephem}).

\begin{table*}
\caption{Minima timings for the eclipsing binary (1+2) in V505~Sgr.
Epoch and $O-C$ were calculated using the ephemeris
${\rm Pri.Min.} = 2433490.483 + 1\fd1828688 \times E$.
$\sigma_{\rm lite}$ denotes assumed standard uncertainty of the minimum determination.
Only newer minima after Chambliss et al. (1993) are listed.
The references are:
Rovithis-Livaniou \& Rovithis (1992),
M\"uyesseroglu et al. (1996),
Ibanoglu et al. (2000),
Cook et al. (2005),
Chochol et al. (2006),
Zasche et al. (2009).
The last 4~measurements are new.
}
\label{tab:lite}
\centering
\begin{tabular}{rrccl}
\tableline\tableline
$JD-2400000$ & Epoch & $O-C$ / d & $\sigma_{\rm lite}$ / d & source\\
\tableline
48432.4871 & 12632.0 & $+0.0054$ &  0.0007 & R.-L. \\
48501.0981 & 12690.0 & $+0.0100$ &  0.0021 & Chochol \\
48858.3253 & 12992.0 & $+0.0109$ &  0.0007 & M\"uyesseroglu \\
51000.4948 & 14803.0 & $+0.0049$ &  0.0007 & Ibanoglu \\
51051.3578 & 14846.0 & $+0.0046$ &  0.0007 & '' \\
51057.2724 & 14851.0 & $+0.0048$ &  0.0007 & '' \\
51064.3692 & 14857.0 & $+0.0044$ &  0.0007 & '' \\
52754.6756 & 16286.0 & $-0.0087$ &  0.0007 & Chochol \\
52843.3891 & 16361.0 & $-0.0103$ &  0.0007 & '' \\
53263.3029 & 16716.0 & $-0.0150$ &  0.0007 & '' \\
53525.8969 & 16938.0 & $-0.0178$ &  0.0007 & Cook \\
53626.4399 & 17023.0 & $-0.0187$ &  0.0007 & Chochol \\
54267.5469 & 17565.0 & $-0.0266$ &  0.0007 & Zasche \\
54267.5472 & 17565.0 & $-0.0263$ &  0.0007 & '' \\
54648.4260 & 17887.0 & $-0.0313$ &  0.0005 & '' \\
54655.5233 & 17893.0 & $-0.0312$ &  0.0003 & '' \\
54658.4817 & 17895.5 & $-0.0299$ &  0.0005 & '' \\
54706.3869 & 17936.0 & $-0.0309$ &  0.0002 & '' \\
55027.5302 & 18207.5 & $-0.0365$ &  0.0018 & Uhl\'a\v r \\
55049.4152 & 18226.0 & $-0.0346$ &  0.0002 & '' \\
55062.4266 & 18237.0 & $-0.0347$ &  0.0011 & \v Smelcer \\
55068.3400 & 18242.0 & $-0.0357$ &  0.0011 & Uhl\'a\v r \\
\tableline 
\end{tabular}
\end{table*}


\subsection{Radial velocities}

We use radial-velocity data from Tomkin~(1992), Tab.~4, who measured 
sharp spectral lines in the 5580 -- 5610\,\AA\ region and attributed
them to the 3rd component. The values of $v_{\rm rad3}$  range from $-13$~to $-9\,{\rm km}/{\rm s}$. 
The width of the lines corresponds to rotational velocity about
$v_{\rm rot3} = (20\pm5)\,{\rm km}/{\rm s}$.

The uncertainties of the radial-velocity data $\sigma_{\rm rv} = 2\,{\rm km}\,{\rm s}^{-1}$
were estimated from a scatter of the RV measurements close in time
and verified by a practical test:
we computed a synthetic spectrum with the same resolution
as Tomkin (1992) and fitted the lines in question by a Gaussian function.
We also checked for possible blends with nearby faint lines.

Wide lines in the V505 Sgr spectrum are attributed to the components
of the eclipsing pair (1+2). The binary is tight and in all likelihood rotates synchronously,
thus the corresponding rotational Doppler broadening is large ($v_{\rm rot1+2} = (100\pm 10)\,{\rm km}/{\rm s}$).
The systemic radial velocity of the (1+2)-body is $v_{\rm rad1+2} = (1.9\pm1.4)\,{\rm km}/{\rm s}$.

\section{Numerical integrator and $\chi^2$ metric}\label{sec:integrator}

In order to model orbital evolution of the multiple-star system V505~Sgr,
namely mutual gravitational interactions of all bodies, we use 
a Bulirsch-St\"oer $N$-body numerical integrator from the SWIFT package
(Levison \& Duncan 1994).

Our method is quite general --- we can model classical Keplerian
orbits, of course, but also non-Keplerian ones (involving 3-body interactions).
We are able to search for both bound (elliptical) and unbound (hyperbolic) trajectories.
Free parameters of our model are listed in Table~\ref{tab:free}.
Hereinafter, we strictly denote individual bodies by numbers:
1, 2 (Algol-type pair), 3 (resolved third component) and 4 to avoid any confusion.

Fixed (assumed) parameters are listed in Table~\ref{tab:dependent}.
Masses of the first three components are well constrained by
photometry and spectroscopy:
$m_1 = (2.20 \pm 0.09)\, M_\odot$,
$m_2 = (1.15 \pm 0.05)\, M_\odot$,
$m_3 = (1.2 \pm 0.1)\, M_\odot$
(Chambliss et al. 1993, Tomkin 1992).
We take $m_3$ as a free parameter, thought,
because of larger relative uncertainty.
When we test 3-body configurations, we have simply $m_4 = 0$.

\begin{table*}
\caption{Free parameters of our dynamical 4-body model.}
\label{tab:free}
\centering
\begin{tabular}{rll}
\tableline\tableline
no. & parameter & brief description			\\
\tableline
 1. & $d$	& distance of the V505 Sgr barycentre	\\

 2. & $m_3$	& mass of the 3rd body			\\
 3. & $z_{h3}$	& position, (1+2)-centric, epoch $T_0$	\\
 4. & $v_{xh3}$	& velocities, (1+2)-centric, epoch $T_0$\\
 5. & $v_{yh3}$	& 					\\
 6. & $v_{zh3}$	& 					\\
 7. & $m_4$	& mass of the 4th body			\\
 8. & $x_{h4}$	& positions, (1+2)-centric, epoch $T_0$	\\
 9. & $y_{h4}$	& 					\\
10. & $z_{h4}$	& 					\\
11. & $v_{xh4}$	& velocities, (1+2)-centric, epoch $T_0$\\
12. & $v_{yh4}$	& 					\\
13. & $v_{zh4}$	& 					\\
\tableline
\end{tabular}
\end{table*}

\begin{table*}
\caption{Fixed (assumed) parameters of our model.}
\label{tab:dependent}
\centering
\begin{tabular}{rll}
\tableline\tableline
no. & parameter & brief description \\
\tableline
14. & $m_{1+2} = 3.4\,M_\odot$	& mass of the (1+2) body \\
16. & $x_{h1+2} = 0\,{\rm AU}$	& positions of the (1+2) body, \\
17. & $y_{h1+2} = 0$	& (1+2)-centric			\\
18. & $z_{h1+2} = 0$	& 				\\
19. & $v_{xh1+2} = 0\,{\rm AU}/{\rm day}$	& velocities			\\
20. & $v_{yh1+2} = 0$	& 				\\
21. & $v_{zh1+2} = 0$	& 				\\
22. & $x_{h3}$	& positions of the 3rd body,		\\
23. & $y_{h3}$	& (1+2)-centric 			\\
24. & $T_0 = 2446282.24375\,{\rm JD}$	& UTC time corresponding \\
    &  (or $2447607.5185$)		& to initial conditions \\
\tableline
\end{tabular}
\end{table*}

\begin{figure*}
\centering
\includegraphics[height=6.2cm]{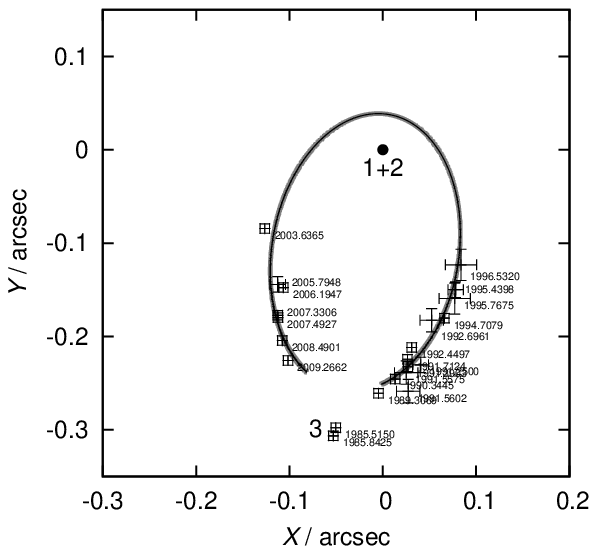}
\includegraphics[height=6.2cm]{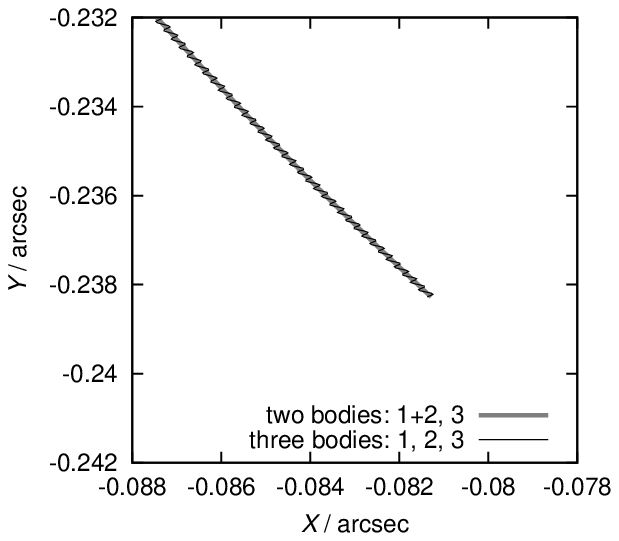}
\caption{Comparison of two 3rd-body trajectories,
computed for three-body (1, 2, 3) and two-body (1+2, 3) configurations.
Left: an overview of the trajectories in a 1-centric frame.
Right: a detail of the small part of the trajectory, where the difference is visible.
Error bars denote speckle-interferometry observations.}
\label{fit_pozorovani_2BODY_V505_Sgr_arcsec}
\end{figure*}

First, it is often useful to adopt a simplification: 1st and 2nd body can be regarded as a single (1+2) body in our dynamical model.
The central pair (1+2) is so compact ($a = 0.033\,{\rm AU}$) and the distance of other components
so large, that it behaves like a single body; its equivalent $J_2$ gravitational moment
is negligible. Indeed, at distance $r = 10\,{\rm AU}$
\begin{equation}
J_2 \simeq {1\over 2} \left({a\over r}\right)^2 {m_1 m_2\over (m_1+m_2)^2} \simeq 10^{-6}\,.
\end{equation}
This can be confirmed easily by a direct numerical integration.
The difference between trajectories computed for three-body (1, 2, 3)
and two-body (1+2, 3) configurations is insignificant and always smaller
than observational uncertainties (see Figure~\ref{fit_pozorovani_2BODY_V505_Sgr_arcsec}).

We also make use of the following two constraints:
(i)~initial positions $x_{h3}$, $y_{h3}$ and zero time $T_0$ of the 3rd body correspond
to a selected speckle-interferometry datum (e.g., the mean of the first two points, or to the third point);
(ii)~3rd body initial velocity components are almost tangent to the observed interferometric
trajectory in the $(x,y)$ plane.

Initial conditions of the integration are specified in an arbitrary
(usually 1+2-centric) frame. We then perform a transformation to
a barycentric frame. The numerical integration runs in the barycentric
Cartesian frame, where
$x, y$~axes correspond to sky plane,
$z$~axis is oriented from the observer towards the system.
We use AU, AU/day units for positions and velocities.

We integrate the system forward for 10,000~days,
and backward (i.e., with opposite sign of initial velocities) for 20,000~days 
in order to cover the observational time span.
The time step used is $\Delta t = 10$~days
and the precision parameter of the BS integrator is $\epsilon = 10^{-8}$.
Finally, we transform the output back to the (1+2)-centric frame
and linearly interpolate the output data to the exact times of observations.


In order to compare the observations to our model
we constructed a~$\chi^2$ metric as follows:
\begin{equation}
\chi^2 = \chi^2_{\rm sky} + \chi^2_{\rm lite} + \chi^2_{\rm rv}\,,\label{eq:chi2}
\end{equation}
where
\begin{equation}
\chi^2_{\rm sky} = \sum_{i=1}^{N_{\rm sky}} {\left(x_{h3}' - x_{h3}[i]\right)^2 + \left(y_{h3}' - y_{h3}[i]\right)^2 \over \sigma_{\rm sky}^2[i]}\,.
\end{equation}
We denote $x_{h3}'$, $y_{h3}'$ (1+2)-centric coordinates of the 3rd body calculated from our model,
which were linearly interpolated to the times $t_{\rm sky}[i]$ of observations $x_{h3}[i]$, $y_{h3}[i]$.
Distance~$d$ is used to convert angular coordinates to AU. Secondly,
\begin{equation}
\chi^2_{\rm lite} = \sum_{i=1}^{N_{\rm lite}} {\left(z_{b1+2}' - z_{b1+2}[i]\right)^2 \over \sigma_{\rm lite}^2[i]}\,,
\end{equation}
where $z_{b1+2}'$ are barycentric coordinates of the (1+2) body computed from our model
and interpolated to the times $t_{\rm lite}[i]$ of observations $z_{b1+2}[i]$.
Because of freedom in the period determination and freedom in the selection of initial velocities,
we have to detrend the light-time effect data (by two LSM fits of $z_{b1+2}'(t)$ and $z_{b1+2}(t)$).
Finally,
\begin{equation}
\chi^2_{\rm rv} = \sum_{i=1}^{N_{\rm rv}} {\left(v_{zh3}'-v_{zh3}[i]\right)^2 \over \sigma_{\rm rv}^2[i]}\,,
\end{equation}
where we again interpolate our model to the times $t_{\rm rv}[i]$.
Note that in case of a 4-body configuration we will attribute the velocities
to the 4th body and change this metric correspondingly (see below).

Optionally, we can add an {\em artificial\/} function to $\chi^2$ in order to constrain the mass~$m_4$
within reasonable limits, e.g.,
\begin{equation}
\chi^2_{m_4} = \left[ \left(m_4 - {m_{4\rm min} + m_{4\rm max}\over 2}\right) \cdot {2\over m_{4\rm max}-m_{4\rm min}} \right]^{100}\,,\label{eq:mass_limit}
\end{equation}
with
$m_{4\rm min} = 0.1\,M_\odot$,
$m_{4\rm max} = 1.2\,M_\odot$.
The upper limit follows from the fact that no other bright star is observed
in the vicinity of V505 Sgr.

A similar expression can be used to constrain the absolute value of velocity $v_4$
(e.g., to be smaller than the escape velocity from the system,
otherwise, we often obtain hyperbolic velocities).

Occasionally, we use a different metric instead of~(\ref{eq:chi2}):
\begin{equation}
\chi^2 = w_{\rm sky} \chi^2_{\rm sky} + w_{\rm lite} \chi^2_{\rm lite} + w_{\rm rv} \chi^2_{\rm rv}\label{eq:chi2_weights}
\end{equation}
with weights $w_{\rm sky} \ge w_{\rm lite}, w_{\rm rv}$, in order to fit the
interferometric trajectory better. There are only five points after
the periastron passage, which would otherwise have too low statistical
significance compared to a lot of light-time data.


What can we expect about the 13-dimensional function $\chi^2(d, m_3, z_{h3}, \dots, v_{zh4})$?
It will surely have many local minima, which would be statistically almost equivalent.
(One can shoot the 4th body from a slightly different position with a slightly different
velocity to get almost the same result.) The problem is degenerate in this sense.
Clearly, there are strong correlations, e.g., between the mass $m_4$ and the
minimal distance of a close encounter (and consequently initial positions/velocities
of the 4th body). Minimisation of the $\chi^2$ function is thus a difficult task.

We use a simplex algorithm (Press et al. 1997) to save computational
resources and to find local minima. However, it is not our goal to find a
global minimum of $\chi^2$, because of the degeneracy and the immense size
of the parameter space. We anyway do not expect a deep, statistically significant
global minimum. Instead, we will choose a set of starting points for the 4th body and
look for a subset of {\em allowed\/} solutions.

On the other hand, in case we test a 3-body configuration only,
the problem is much simpler: the six-dimensional
$\chi^2(d, m_3, z_{h3}, v_{xh3}, v_{yh3}, v_{zh3})$
is well-behaved and we may expect to find a unique solution
(and its uncertainty).


\section{Results}\label{sec:results}

In the following subsections we consider and analyse several
hypotheses about the nature of the V505 Sgr system:
  (i)~there are three bodies only in V505 Sgr;
 (ii)~the 3rd body directly perturbs the central pair;
(iii)~a steady mass transfer causes minima timing variations;
 (iv)~there is modulation of mass transfer by the 3rd body;
  (v)~a sudden mass transfer occurred around 2000;
 (vi)~Appelgate's mechanism is operating;
(vii)~a 4th body is present (either on a bound or hyperbolic orbit).


\subsection{The 3rd body alone on a Keplerian orbit}\label{sec:keplerian_orbit}

\begin{figure*}
\centering
\includegraphics[height=4.7cm]{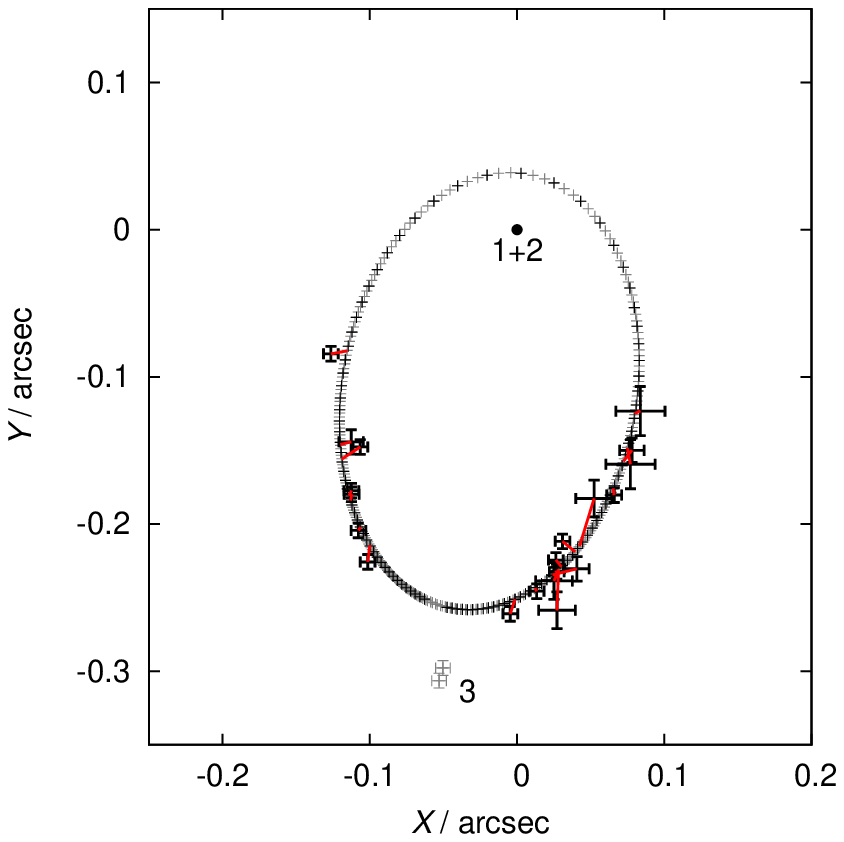}
\includegraphics[height=4.7cm]{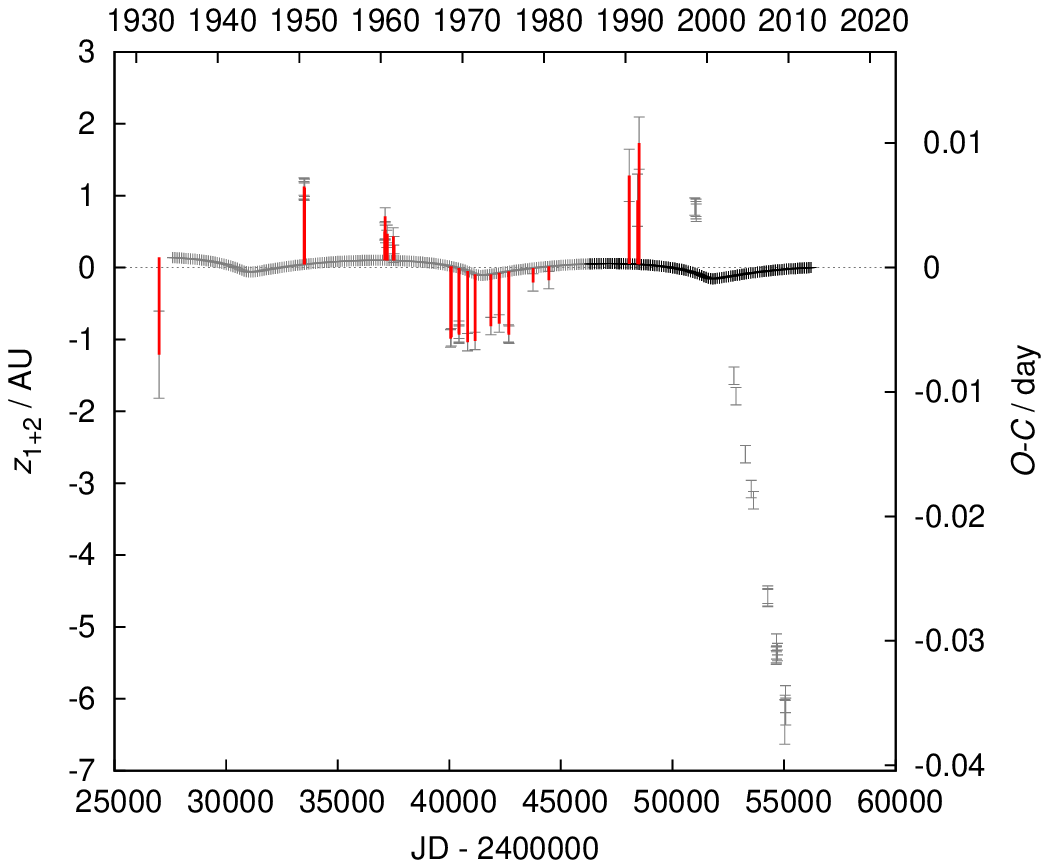}
\includegraphics[height=4.7cm]{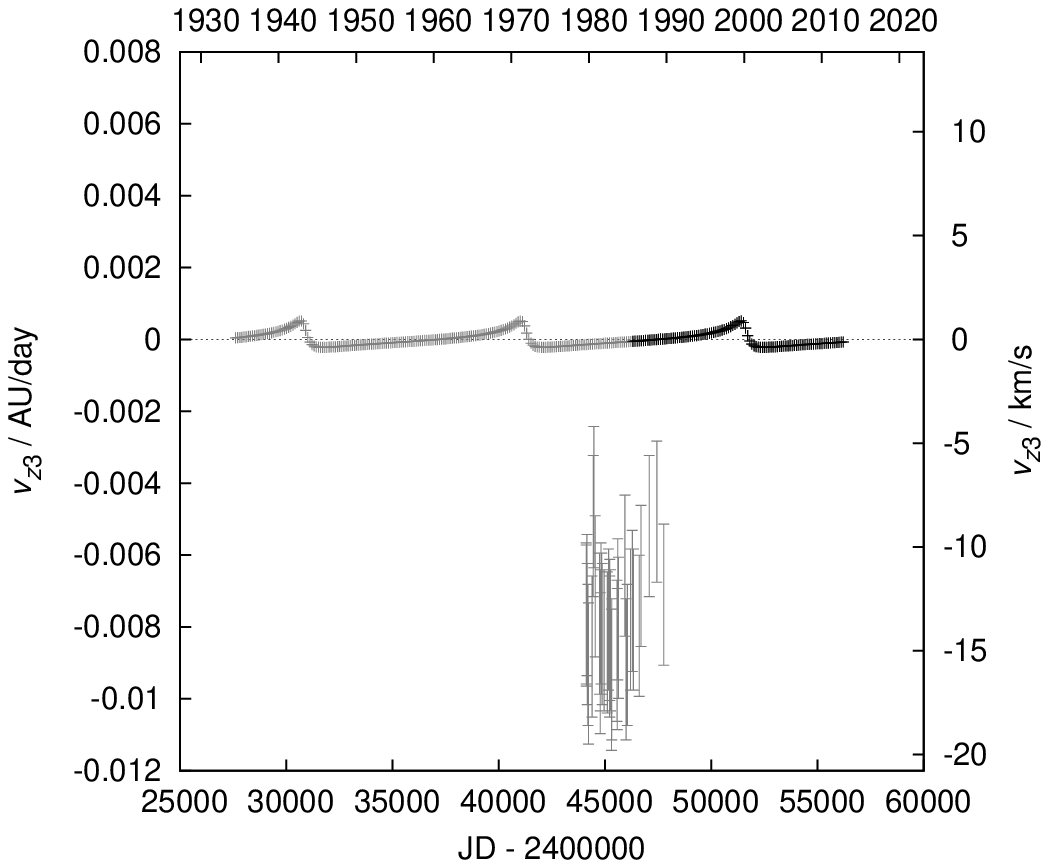}
\caption{Best-fit solution for the trajectory of the 3rd body, which corresponds
to speckle interferometry data (but excluding 1985 measurements).
Neither light-time effect nor radial velocities were fitted in this this case.
Parameters of the 3rd body are:
$m_3 = 1.17\,M_\odot$,
$x_{h3} = -0.15\,{\rm AU}$,
$y_{h3} = -25.9\,{\rm AU}$,
$z_{h3} = -0.63\,{\rm AU}$,
$v_{xh3} = 0.0037\,{\rm AU}/{\rm day}$,
$v_{yh3} = 0.0017\,{\rm AU}/{\rm day}$,
$v_{zh3} = 0.0000\,{\rm AU}/{\rm day}$
for $T_0 = 2447607.5185\,{\rm JD}$.
The inclination of the orbit is very low in this case ($I = 1.5^\circ$).
The resulting $\chi^2_{\rm sky} = 52$, with the number of data points $N_{\rm sky} = 20$.
Note there is a strong disagreement of this Keplerian orbit with
both $O-C$ data and radial velocities (total $\chi^2 = 1700$, $N = 90$).}
\label{simplex3rd_VZDALENOST2_WO_LITE_V505_Sgr_arcsec}
\end{figure*}

At first, let us test a standard "null" hypothesis, i.e.,
only 3rd body exists ($m_4 = 0$). It is possible to fit speckle data {\em alone\/}
($w_{\rm lite} = w_{\rm rv} = 0$)
by an elliptical orbit with a $(29\pm 1)$-yr period, especially,
if we assume the first two 1985 measurements are erroneous
(offset by 50\,mas, see Figure~\ref{simplex3rd_VZDALENOST2_WO_LITE_V505_Sgr_arcsec}, left).
The $\chi^2_{\rm sky} = 50$ for this fit and the respective number of data points is $N_{\rm sky} = 20$.
(Thought ideally, $\chi^2$ should be comparable to $N$.)

Note the $\chi^2_{\rm sky}$ would be much higher, if we include the 1985 measurements:
$\chi^2_{\rm sky} = 210$, $N_{\rm sky} = 22$. It means, if these two measurements
are not systematic errors, the 29-yr Keplerian orbit is essentially excluded!
The two respective measurements were obtained by two {\em different\/} telescopes during
two different nights (see McAlister et al. (1987a) and McAlister et al. (1987b)).
We checked measurements of another 34 stars in these publications, observed
with the {\em same\/} telescope and during the same night as V505 Sgr, and
we have found no indication of a wrong plate scale --- all measurements lie
on Keplerian ellipses within usual observational uncertainties (5\,mas).
We thus belive the 1985 measurements are {\em not\/} erroneous and they
should be included in the $\chi^2$ metric.

Without additional (non-positional) data it is not possible to distinguish between different inclinations
--- there are equivalent low-$I$ and high-$I$ solutions with almost the same $\chi^2 \simeq 50$.
Nevertheless, {\em every\/} inclined orbit of the 3rd body has to cause
a corresponding light-time effect, otherwise must be considered wrong!
Even a slight $I \gtrsim 2^\circ$ inclination would be easily detectable
in the light-time effect data (see Figure~\ref{simplex3rd_VZDALENOST2_WO_LITE_V505_Sgr_arcsec}, middle).
A period analysis of the $O-C$ data (with Period04 program) also does not show
a prominent 29-yr period. On the other hand, there is a clear signal at
$P = 39\,{\rm yr}$, with an amplitude of the peak $A = 0.0092\,{\rm d}$.

If we assume the $O-C$ data are indeed caused by a light-time effect,
there is a strong disagreement of the 29-yr Keplerian orbit with the light-time
effect data (and also with radial velocities), even prior to 2000!
If we try to fit the whole orbit and light-time effect data {\em together\/},
we would have $\chi^2_{\rm sky} = 107$ and $N_{\rm sky} = 20$, i.e., such an orbit
is excluded with a high significance. There are also clear systematic departures
between the observed interferometric data and calculated Keplerian orbit.

The only possibility is the inclination of the 3rd-body orbit is almost
zero $I \lesssim 2^\circ$, so we do not see any light-time effect at all.
The observed $O-C$ variations then must caused by an entirely different phenomenon
(see next Sections~\ref{sec:period_change} to \ref{sec:applegate_1992} for a detailed discussion).

Nevertheless, there still remains a strong disagreement with the observed
high radial velocities $v_{{\rm rad}3} \simeq 10\,{\rm km}/{\rm s}$,
because a non-inclined orbit should have $v_{{\rm rad}3} \lesssim 1\,{\rm km}/{\rm s}$.
We have no solution for this problem (unless there is a 4th body present
in the system, see Sections~\ref{sec:optimalizace_VZDALENOST4} to \ref{sec:simplex_4TH_BODY9}).

\subsection{Direct perturbation of the 1+2 orbital period by the 3rd body}\label{sec:period_change}

\begin{figure}
\centering
\includegraphics[width=7.5cm]{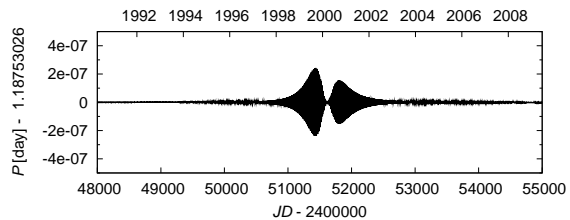}
\caption{Simulated osculating orbital period~$P$ of the central binary (bodies 1 and 2),
perturbed by the 3rd body. The periastron passage occurred in 2000
and the corresponding change of period is $\Delta P \lesssim 10^{-7}\,{\rm d}$.
The observed values of $|\Delta P| \simeq 10^{-5}\,{\rm d} $ are much larger
than in this simulation.}
\label{Pt}
\end{figure}

One may ask, if the observed variations in minima timings, which correspond to
the changes of the period of the order $|\Delta P| \simeq 10^{-5}\,{\rm d}$,
could be caused by a {\em direct\/} gravitational perturbation of the tight central pair (1, 2)
by the orbiting 3rd body. In periastron, the minimum distance is of the order
$\simeq 10\,{\rm AU}$.
In order to test this possibility, we use our dynamical model with three
bodies 1, 2 and 3 taken separately. A detection of minute changes of the orbital period
requires a smaller time step and higher precision of the BS integrator ($\Delta t = 0.01\,{\rm d}$, $\epsilon = 10^{-12}$).
The resulting osculating orbital period changes during one periastron passage
are shown in Figure~\ref{Pt}. They are much smaller than $\Delta P \lesssim 10^{-7}\,{\rm d}$.
An extremely close encounter (within less then 0.1\,AU, which corresponds to 0.001\,arcsec)
would be needed to change the orbital period of the tight Algol system substantially.

Moreover, {\em anything\/} directly connected with the 3rd body should conform
to the 39~year period of the minima timings and this, according to Section~\ref{sec:keplerian_orbit},
is in conflict with any 29-yr Keplerian orbit of the 3rd body.

\subsection{Effects of mass transfer between 1 and 2}\label{sec:mass_transfer}

\begin{figure}
\centering
\includegraphics[width=7.5cm]{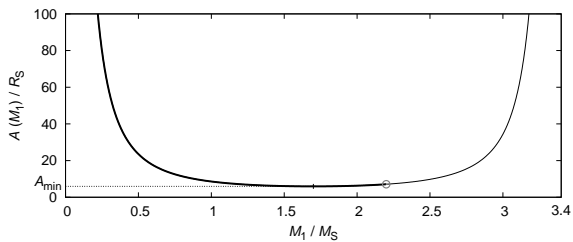}
\caption{The dependence of separation~$A$ of the eclipsing binary on the mass~$M_1$ of the 1st body,
resulting from the conservation of total mass and orbital angular momentum.}
\label{vzdalenost_V505Sgr}
\end{figure}

Past photometric and spectroscopic observations confirm
the central pair of V505 Sgr is a classical semi-detached Algol system,
with a less-massive secondary filling its Roche lobe (Chambliss et al. 1993).
In case of a conservative mass transfer, the sum of masses is constant
\begin{equation}
M_1(t) + M_2(t) = K\,,\label{eq:ZZHM}
\end{equation}
as well as the orbital angular momentum
\begin{equation}
A(t) M_1^2(t) M_2^2(t) = C\,,\label{eq:ZZMH}
\end{equation}
where $A(t)$ denotes the actual separation of the stars.
We can substitute current masses and separation $A = 7.1\,R_\odot$ (Chambliss et al. 1993)
into these equations, compute constants $K$, $C$ and consequently the dependence
$A(M_1)$ (see also Figure~\ref{vzdalenost_V505Sgr})
\begin{equation}
A(M_1) = C M_1^{-2} (K-M_1)^{-2}\,.\label{eq:A_M_1}
\end{equation}
A smooth conservative mass transfer should increase orbital period steadily,
since in the V505 Sgr case the mass ratio has been reversed already ($M_1 > M_2$).
On contrary, we observe an abrupt {\em decrease\/} of the period
$\Delta P = -1.2 \times 10^{-5}\,{\rm d}$ after 2000.
We thus conclude a simple mass transfer cannot explain
the observer minima timings.

\subsection{Modulation of mass transfer between 1 and 2 during the 3rd body encounter}\label{sec:roche}

In this section, we test if the 3rd body is capable to change the Roche
potential of the central binary (bodies 1 and 2) in a such a way, that
the mass transfer rate ${\rm d} M/{\rm d} t$ (and consequently ${\rm d} P/{\rm d} t$)
changes by a substantial amount.
We add a 3rd-body term to the Roche potential
\begin{equation}
\Omega(x,y,z) = {1\over r_1} + {q\over r_2} + {1\over 2} (1+q) r_3^2 + {q_{\rm 3rd}\over r_{\rm 3rd}}\,,\label{eq:roche_3rd}
\end{equation}
where $q = M_2/M_1$ denotes the mass ratio and similarly $q_{\rm 3rd} = M_3/M_1$.
We see immediately, that relative change of the potential due to the 3rd body at distance $r_{\rm 3rd} \simeq 10\,{\rm AU}$ is
$\delta\Omega/\Omega \simeq 10^{-13}$. We do not find it likely, that such
a minuscule perturbation of the potential, and thus the related tidal acceleration,
could produce significant effects. Consequently, we cannot explain minima
timings variations by the modulation of mass transfer.
Finally, similarly as in Section~\ref{sec:period_change}, this effect
would be also in conflict with a 29-yr Keplerian orbit of the 3rd body.


\subsection{A sudden mass transfer of Biermann \& Hall (1973)}\label{sec:biermann_hall_1973}

According to Biermann \& Hall (1973) a sudden mass transfer between
the Algol components may result in a {\em temporary\/} decrease
of the orbital period, even thought mass is flowing from the lighter
component to the more massive. In our case, we would need ${\rm d}M/{\rm d}t$ as high as $\simeq 10^{-6}\,M_\odot/{\rm yr}$
to explain period changes $|{\rm d} P/{\rm d} t| \simeq 10^{-6}\,{\rm d}/{\rm yr}$.
Such a mass transfer rate seems to be too large
compared to theoretical models (Harmanec 1970),
${\rm d}M/{\rm d}t \gtrsim 10^{-6}\,M_\odot/{\rm yr}$ are reached
only during a very short interval of time, before the reversal
of mass ratio.

Another problem of this scenario is that we observe rather smooth
periodic variations of the minima timings before 2000, which do not
seem to be entirely compatible with this mechanism, which may be more irregular in time. 
This phenomenon is also rarely confirmed by independent observations.
(It would require a very precise photometry on a long time scale,
or a spectroscopic confirmation of circumstellar matter.)
Today, this mechanism is not generally accepted as a major cause
of minima timing variations among Algol-type systems.


\subsection{Applegate (1992) magnetic mechanism}\label{sec:applegate_1992}

Applegate (1992) proposed a gravitational quadrupole coupling of orbit
and shape variations of a magnetically active subgiant (2nd component)
can result in variations of the orbital period and hence minima timings.
In this scenario, the observed 39-yr period would correspond to the period
of the magnetic dynamo.

The 2nd (G5~IV) star rotates quickly (1.2\,d), it has a convective envelope
in this evolutionary stage and, presumably, there is a differential rotation
and operating dynamo, which can result in a sufficiently strong magnetic field ($10^4\,{\rm G}$),
necessary for Applegate's mechanism to work.
Period changes of the order $\Delta P/P \simeq 10^{-5}$ should also correspond
to changes of the luminosity $\Delta L_2/L_2 = 0.1$, in phase with minima timings.
Unfortunately, we are not able to confirm this by our photometry
(0.01\,mag precision over tens of years would be required).

In principle, this mechanism can explain minima timings variations,
but it is not clear, why there is an abrupt change after 2000.
An independent confirmation is rare and difficult. One of the possibilities
might be a spectroscopic observation of magnetically active lines
(Ca~II H and K, or Mg~II).
This scenario also does not provide any solution for the observed
large radial velocities.


\subsection{Distance, mass and the 3rd body orbit (prior to 2000)}\label{sec:optimalizace_VZDALENOST4}

Hereinafter, we {\em assume\/} minima timings variations are caused mainly
by the light-time effect due to the orbiting 3rd body.
Because the orbit of the 3rd body prior the periastron passage in 2000
seems unperturbed, we first determine the optimal distance~$d$
of the system, 3rd-body mass~$m_3$ and orbit ($z_3$, $v_{xh3}$, $v_{yh3}$, $v_{zh3}$).
We use only the observational data older than 2000 for this purpose.

We compute $\chi^2$ values for the following set of initial conditions
(we do not use a simplex here):
$d \in (95, 105)\,{\rm pc}$, $\Delta d = 1\,{\rm pc}$,
$m_3 \in (1.1, 1.3)\,M_\odot$, $\Delta m_3 = 0.1\,M_\odot$,
$z_{h3} \in (2.0, 8.0)\,{\rm AU}$, $\Delta z_{h3} = 1.0\,{\rm AU}$,
$v_{xh3} \in (0.0033, 0.0040)\,{\rm AU}/{\rm day}$,
$v_{yh3} \in (0.0008, 0.0016)\,{\rm AU}/{\rm day}$,
$v_{zh3} \in (-0.0018, 0.0012)$ ${\rm AU}/{\rm day}$,
$\Delta v_{xh3} = \Delta v_{yh3} = \Delta v_{zh3} = 0.0001\,{\rm AU}/{\rm day}$.

\begin{figure*}
\centering
\includegraphics[height=4.9cm]{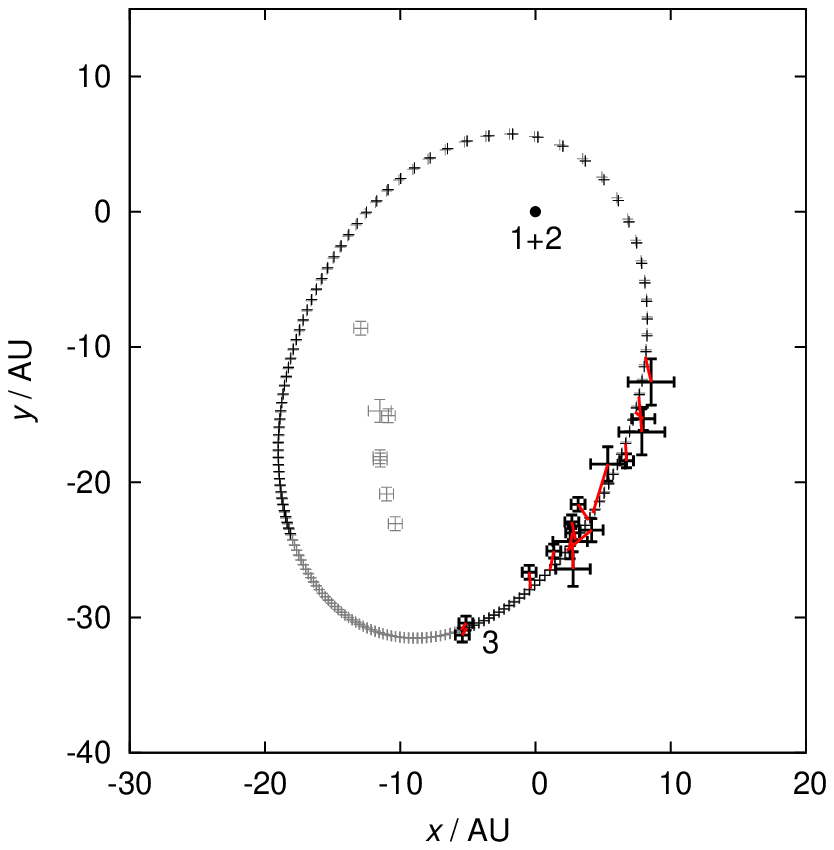}
\includegraphics[height=4.9cm]{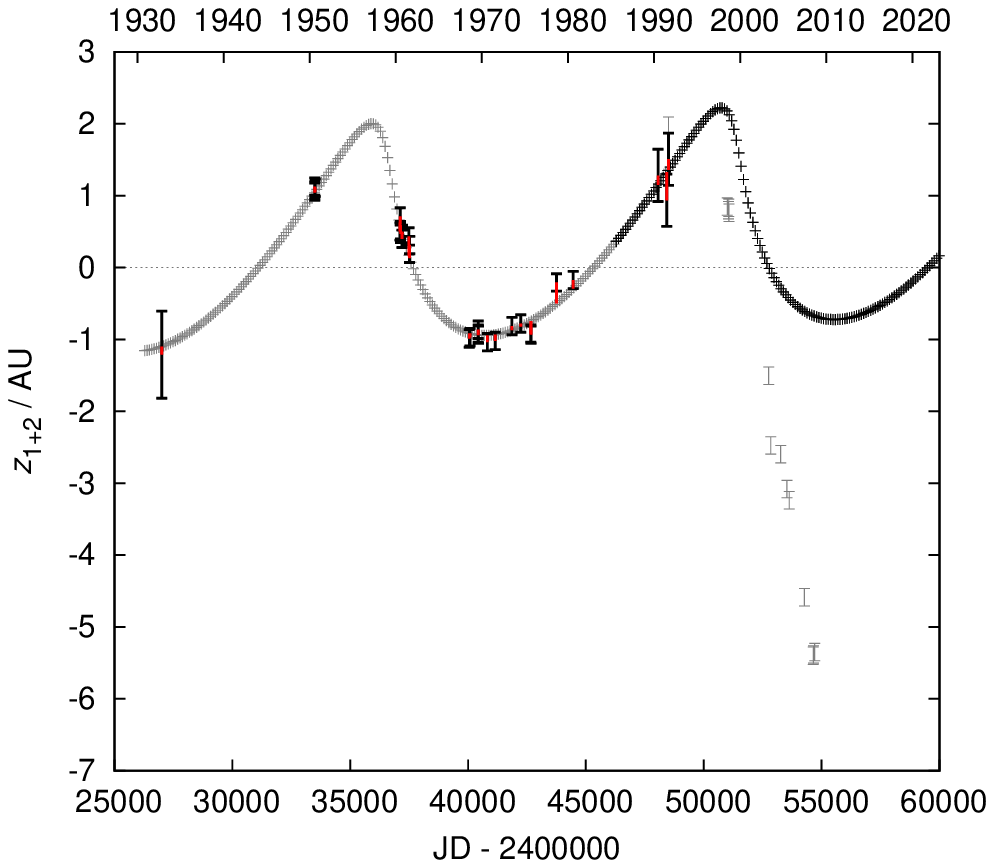}
\includegraphics[height=4.9cm]{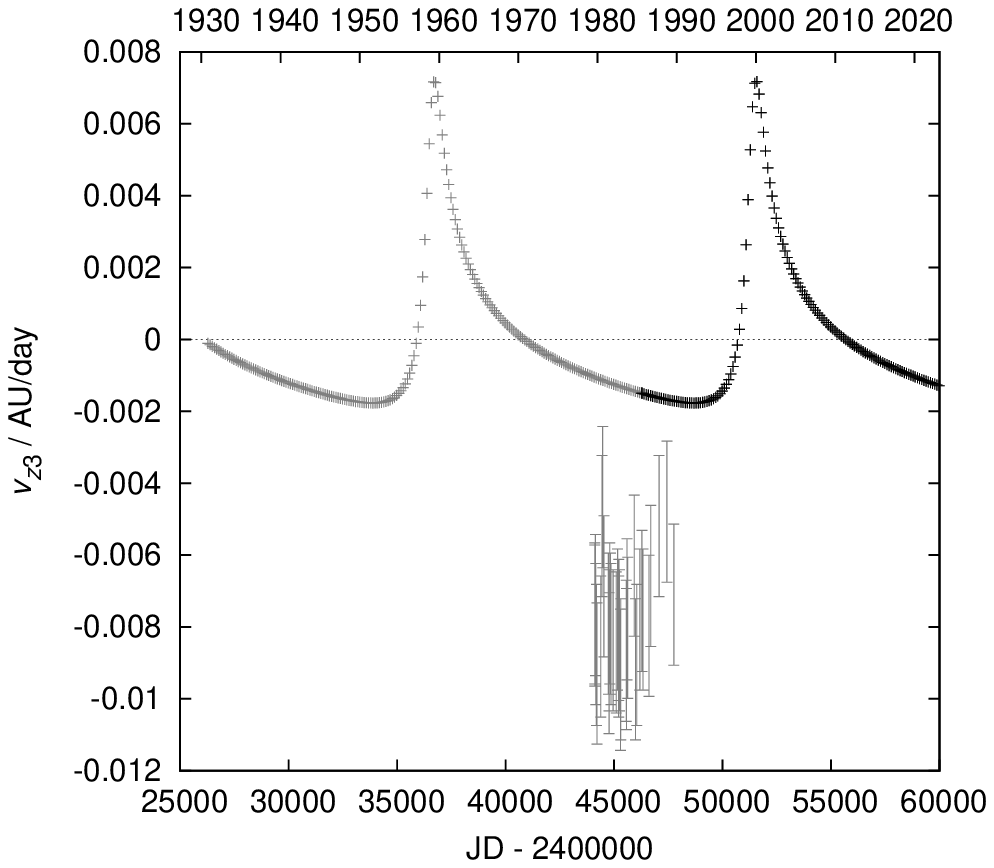}
\caption{Best-fit solution for the orbit of the 3rd body {\em and\/} the light-time effect before 2000:
$d = 102\,{\rm pc}$,
$m_3 = 1.2\,M_\odot$,
$z_{h3} =  4.0\,{\rm AU}$,
$v_{xh3} =  0.0036\,{\rm AU}/{\rm day}$,
$v_{yh3} =  0.0013\,{\rm AU}/{\rm day}$,
$v_{zh3} = -0.0015\,{\rm AU}/{\rm day}$
for $T_0 = 2446282.24375\,{\rm JD}$.
The resulting $\chi^2 = 88$, with the total number of data points $N = 46$.
The red lines denote differences between observed and calculated data.
Radial velocities were not fitted in this case, they are shown for comparison only.}
\label{optimalizace_VZDALENOST4_V505_Sgr_SKY}
\end{figure*}

The best-fit solution is displayed in Figure~\ref{optimalizace_VZDALENOST4_V505_Sgr_SKY}.
Orbital period of the 3rd body is $P = (39\pm 2)\,{\rm yr}$.
The resulting distance $d = (102\pm 5)\,{\rm pc}$. This solution is very similar to that
in Mayer (1997). The parallactic distance of V505~Sgr
given by Hipparcos ($\pi = (8.40\pm0.57)\,{\rm mas}$, $d = 111\hbox{ to }128\,{\rm pc}$, cf., van Leeuwen 2007)
is offset and even the error intervals do not overlap.

Note that the radial velocities of the order $-10\,{\rm km}/{\rm s}$ measured by Tomkin (1992)
cannot be attributed to the 3rd body, which orbital velocity should be much
smaller ($(-2.5\pm0.5)\,{\rm km}/{\rm s}$) according to interferometric and light-time effect data.
Consequently, we do not fit the velocities in this case ($w_{\rm rv} = 0$),
we are going to attribute them to the 4th body (in the next Section~\ref{sec:optimalizace_4TH_BODY_7D_detail}).

Finally, it is important to mention that our solution does {\em not\/} depend
on the two ("offset") 1985 speckle measurements at all! We can exclude them
completely from our considerations and the result would be the same.
Our only assumption was that minima timings variations are caused by the
light-time effect and this enforces the orbital period of $P \simeq 39\,{\rm yr}$.
(But coincidentally, both 1985 measurements fit perfectly this longer-period orbit.)


%

%
%
%
%
%
%

\subsection{Encounter with a 4th body (a $\chi^2$ map)}\label{sec:optimalizace_4TH_BODY_7D_detail}

We next {\em fix\/} initial conditions of the 3rd body according to the results in Section~\ref{sec:optimalizace_VZDALENOST4}
and model a perturbation by a 4th body under different geometries.

The free parameters of the model are:
$m_4$,
$x_{h4}$,
$y_{h4}$,
$z_{h4}$,
$v_{xh4}$,
$v_{yh4}$,
$v_{zh4}$.
We include radial-velocity data, but we assume the spectral lines
(and corresponding velocities) belong to the 4th body.
We scan the following limited set of initial conditions (over 8~million trials):
$m_4 \in (0.5, 0.8)\,M_\odot$, $\Delta m_4 = 0.05\,M_\odot$,
$x_{h4} \in (38, 45)\,{\rm AU}$, $\Delta x_{h4} = 1.0\,{\rm AU}$,
$y_{h4} \in (37, 40)\,{\rm AU}$, $\Delta y_{h4} = 0.5\,{\rm AU}$,
$z_{h4} \in (20, 30)\,{\rm AU}$, $\Delta z_{h4} = 1.0\,{\rm AU}$,
$v_{xh4} \in (-0.011, -0.005)\,{\rm AU}/{\rm day}$,
$v_{yh4} \in (-0.010, -0.005)\,{\rm AU}/{\rm day}$,
$v_{zh4} \in (-0.012, -0.006)$ ${\rm AU}/{\rm day}$,
$\Delta v_{xh4} = \Delta v_{yh4} = \Delta v_{zh4} = 0.0005\,{\rm AU}/{\rm day}$.

\begin{figure*}
\centering
\includegraphics[height=4.8cm]{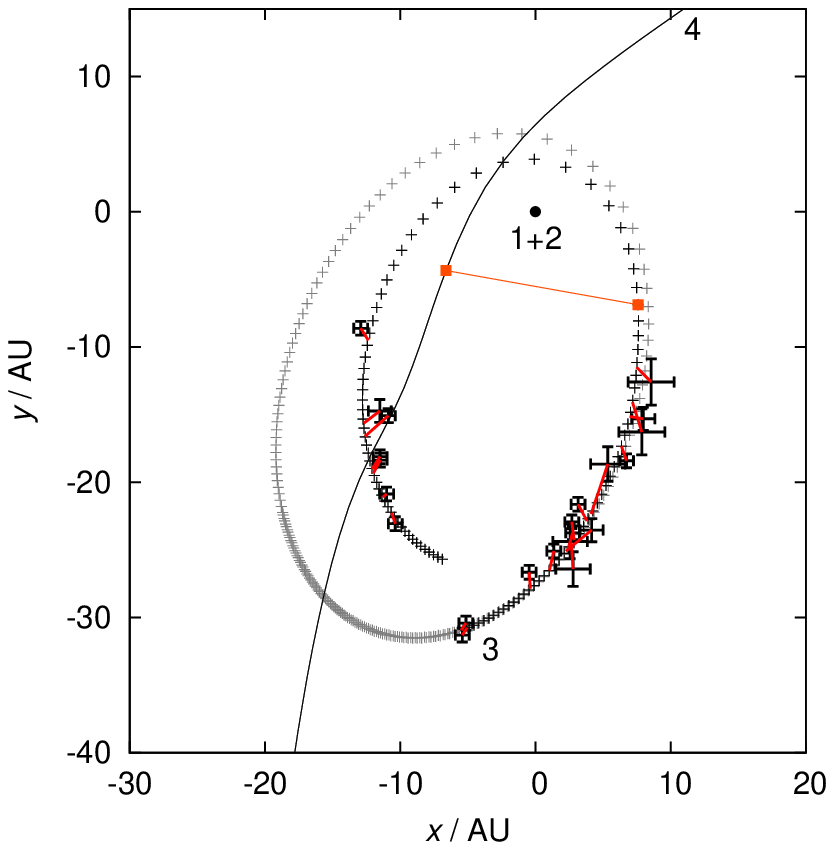}
\includegraphics[height=4.8cm]{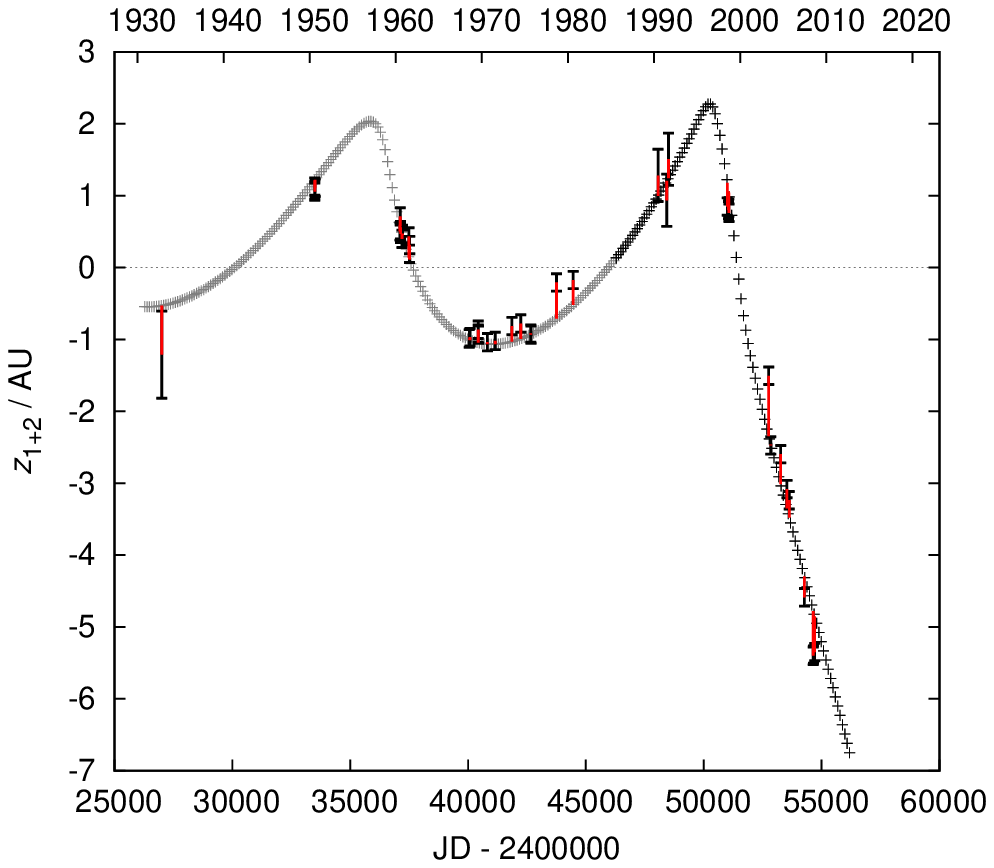}
\includegraphics[height=4.8cm]{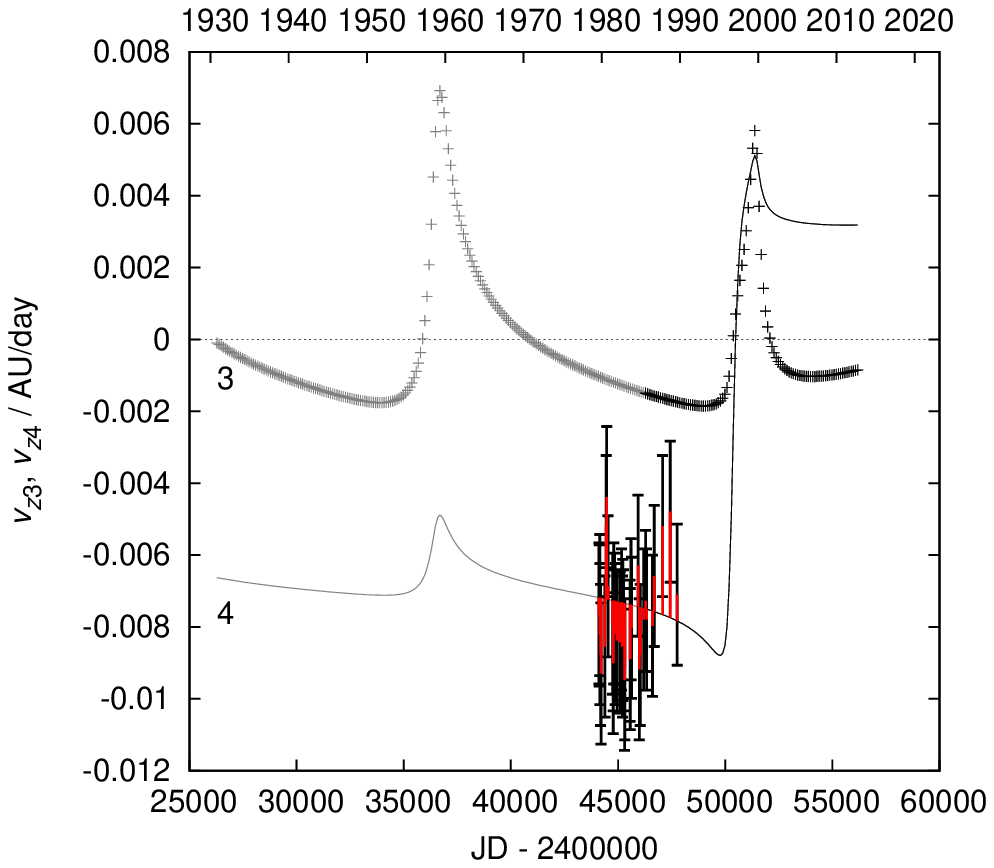}
\caption{Best-fit solution for the trajectory of the 4th body, which best explains
the observed trajectory of the 3rd body, light-time effect and radial velocities:
$m_4 = 0.8\,M_\odot$,
$x_{h4} = 45.0\,{\rm AU}$,
$y_{h4} = 39.5\,{\rm AU}$,
$z_{h4} = 28.0\,{\rm AU}$,
$v_{xh4} = -0.0105\,{\rm AU}/{\rm day}$,
$v_{yh4} = -0.008\,{\rm AU}/{\rm day}$,
$v_{zh4} = -0.0075\,{\rm AU}/{\rm day}$.
The resulting $\chi^2 = 331$, with the total number of data points $N = 102$.
The motion of the 4th body captured in the left panel spans from 1994 to 2003.
The squares connected by a straight line indicate the closest encounter between
the 3rd and 4th body.}
\label{optimalizace_4TH_BODY_7D_detail_V505_Sgr_SKY}
\end{figure*}

\begin{figure}
\centering
\includegraphics[height=5.2cm]{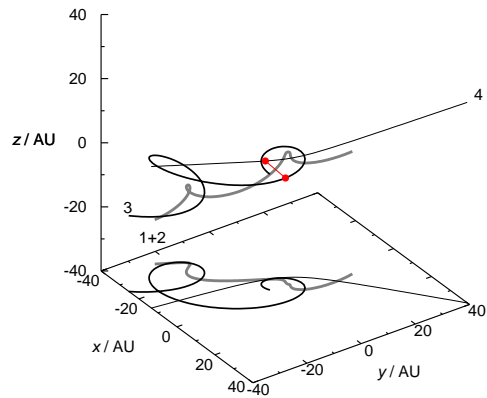}
\caption{Orbits of bodies (1+2), 3, 4 in a barycentric frame and their projection to the plane ($z = -40\,{\rm AU}$).}
\label{optimalizace_4TH_BODY_7D_detail_XYZ_baryc}
\end{figure}

\begin{figure}
\centering
\includegraphics[width=7.5cm]{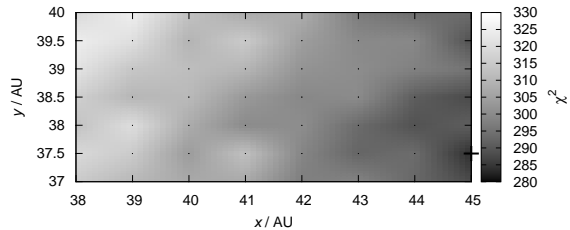}
\caption{Minima of the 7-dimensional function $\chi^2(m_4, x_{h4}, y_{h4}, z_{h4}, v_{xh4},$ $ v_{yh4}, v_{zh4})$
at given positions ($x_{h4}, y_{h4}$). The minimum is taken over remaining free parameters
$m_4, z_{h4}, v_{xh4}, v_{yh4}, v_{zh4}$.
Cross is an overall minimum, black dots represent computed points.}
\label{optimalizace_4TH_BODY_7D_detail_optimalizace_x_y_MIN}
\end{figure}

A comparison of the best-fit solution with observational data is displayed in Figure~\ref{optimalizace_4TH_BODY_7D_detail_V505_Sgr_SKY}.
We use a modified metric~(\ref{eq:chi2_weights}) with $w_{\rm sky} = 10$, $w_{\rm lite} = w_{\rm rv} = 1$.
The respective trajectories of the bodies are shown in Figure~\ref{optimalizace_4TH_BODY_7D_detail_XYZ_baryc}.
Note, however, that according to the $\chi^2$~map (Figure~\ref{optimalizace_4TH_BODY_7D_detail_optimalizace_x_y_MIN})
there are many local minima, which cannot be distinguished from a statistical point of view,
because the values of $\chi^2$ differ only little ($\chi^2 \in [284, 325]$).
The corresponding $\chi^2$ probabilities $Q(\chi^2|N)$, that the observed value of $\chi^2 = 340$
(for a given number of degrees of freedom $N = 105$) is that large by chance even for a correct model,
are too low (essentially zero). It may also indicate that real uncertainties might be a bit larger (by a factor of 2)
than the values estimated by us. Nevertheless, we will find better solutions using a simplex method
(in Section~\ref{sec:simplex_4TH_BODY9}).


%
%
%
%
%
%
%




\subsection{Encounter with a 4th body (different geometry, simplex)}\label{sec:simplex_4TH_BODY9}

We selected a different set of initial conditions for the following modelling.
They serve as starting points for the simplex algorithm:
$m_4 = 0.5\,M_\odot$,
$x_{h4} \in (-100, -10)\,{\rm AU}$,
$y_{h4} \in (-50.1, -0.1)\,{\rm AU}$,
$z_{h4} \in (0, 50)\,{\rm AU}$,
$\Delta x_{h4} = \Delta y_{h4} = \Delta z_{h4} = 5.0\,{\rm AU}$,
$v_{xh4} \in (0.005, 0.015)\,{\rm AU}/{\rm day}$,
$\Delta v_{xh4} = 0.001\,{\rm AU}/{\rm day}$,
$v_{yh4} \in (0, 0.01)\,{\rm AU}/{\rm day}$,
$\Delta v_{yh4} = 0.002\,{\rm AU}/{\rm day}$,
$v_{zh4} \in (-0.007, 0)\,{\rm AU}/{\rm day}$,
$\Delta v_{zh4} = 0.001\,{\rm AU}/{\rm day}$.
The total number of trials reaches $10^6$.

\begin{figure*}
\centering
\includegraphics[height=4.8cm]{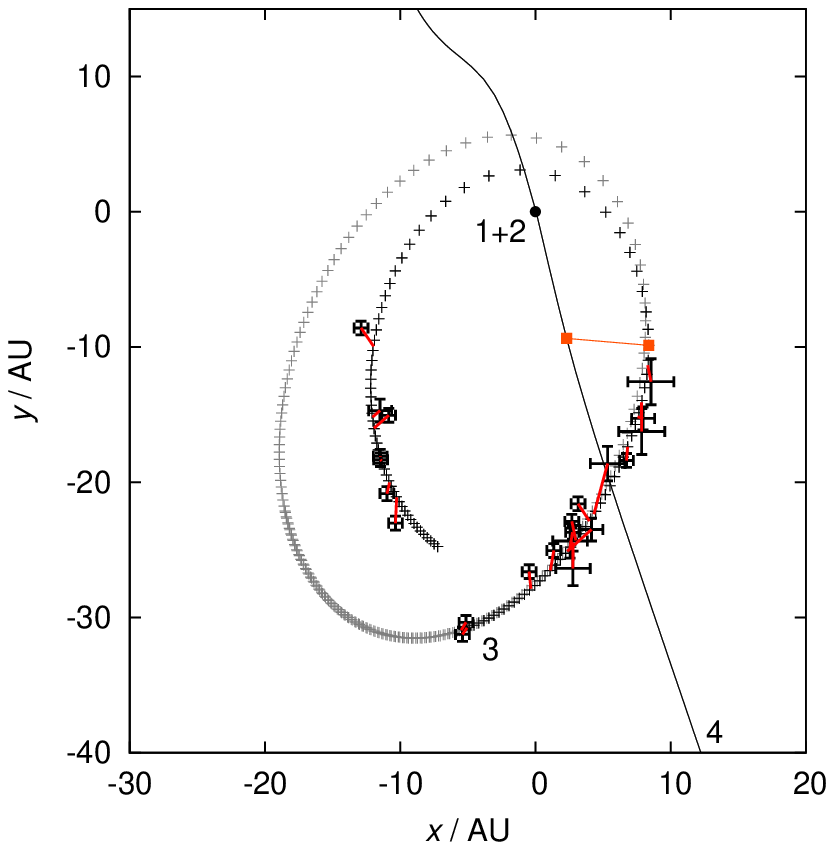}
\includegraphics[height=4.8cm]{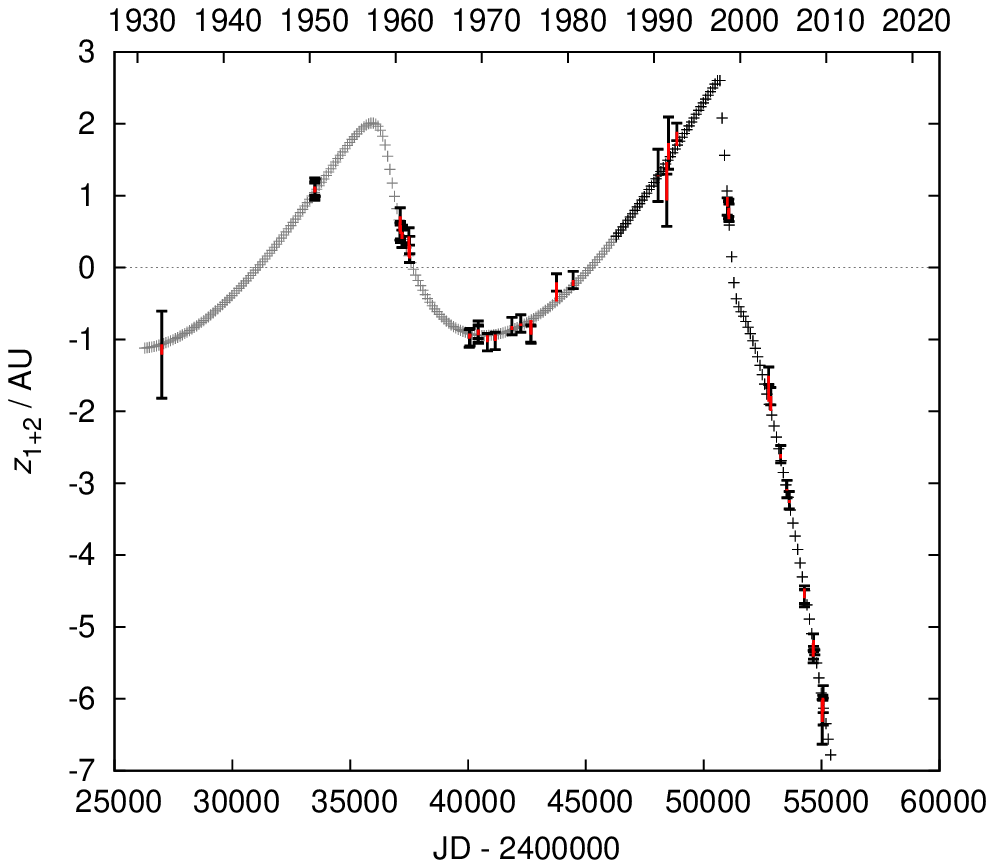}
\includegraphics[height=4.8cm]{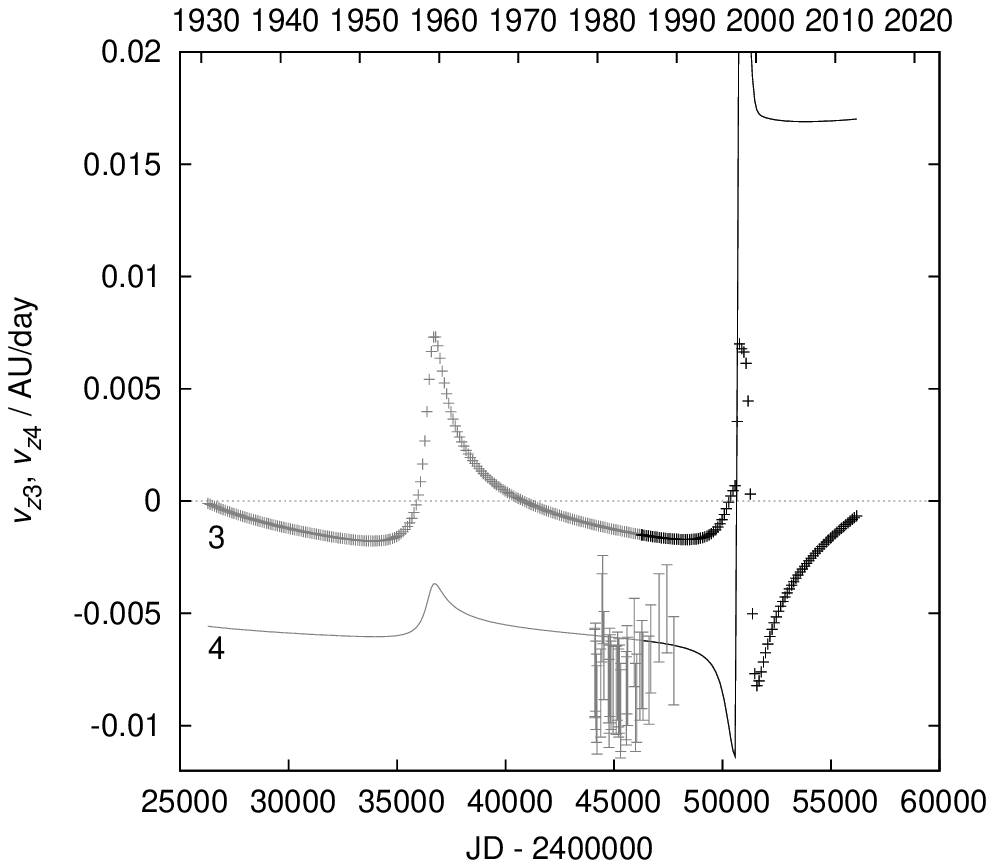}
\caption{A typical solution (out of many) for the trajectory and light-time effect (without radial velocities):
$m_4 = 0.576\,M_\odot$,
$x_{h4} =  27.912\,{\rm AU}$,
$y_{h4} = -84.809\,{\rm AU}$,
$z_{h4} =  30.482\,{\rm AU}$,
$v_{xh4} = -0.00584\,{\rm AU}/{\rm day}$,
$v_{yh4} =  0.01612\,{\rm AU}/{\rm day}$,
$v_{zh4} = -0.00620\,{\rm AU}/{\rm day}$.
The corresponding $\chi^2 = 168$, with the number of data points $N = 73$.
This solution can be further improved by a 15-dimensional simplex
to reach $\chi^2$ as low as 130.}
\label{simplex_4TH_BODY9_V505_Sgr_SKY}
\end{figure*}

We reject radial velocity constraints ($w_{\rm rv} = 0$), although we can find a lot of allowed solutions
with velocities in the correct range ($v_{zh4} \doteq -0.008\,{\rm AU}/{\rm day}$).
On the other hand, we use a mass limit according to Eq.~(\ref{eq:mass_limit}).
An example of a typical good fit is shown in Figure~\ref{simplex_4TH_BODY9_V505_Sgr_SKY}.
We selected one with mass around $m_4 = 0.6\,M_\odot$;
the corresponding $\chi^2 = 168$, $N = 73$ and probability $Q(\chi^2|N) \simeq 10^{-9}$,
still too low.
This solution can be further improved by a 15-dimensional simplex
(i.e., with all parameters of the 3rd body free) to reach $\chi^2$ as low as 130
and $Q(\chi^2|N)$ as high as $10^{-5}$.

\begin{figure*}
\centering
\includegraphics[height=4.8cm]{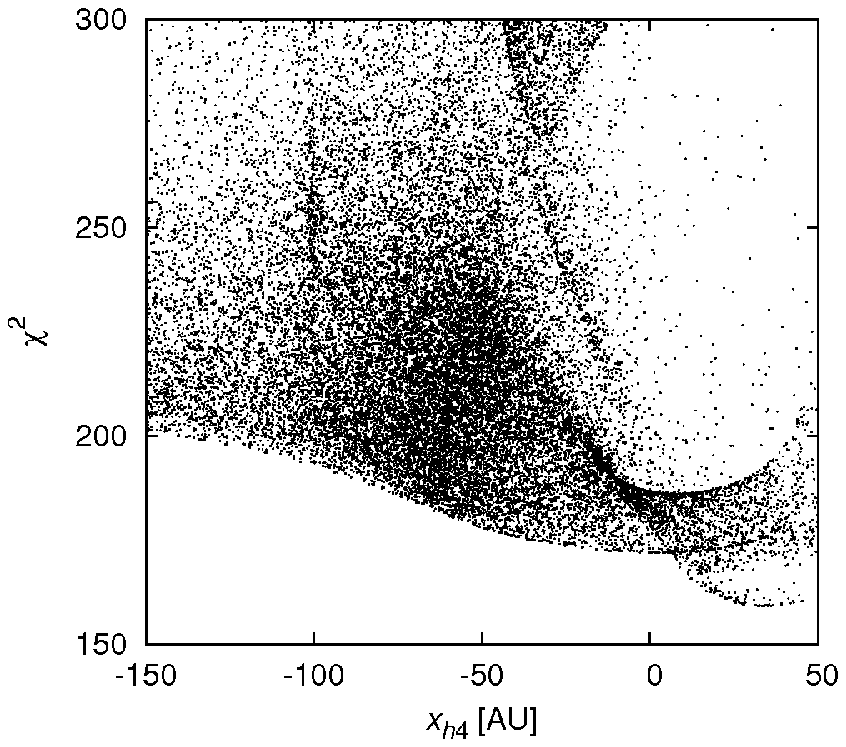}
\includegraphics[height=4.8cm]{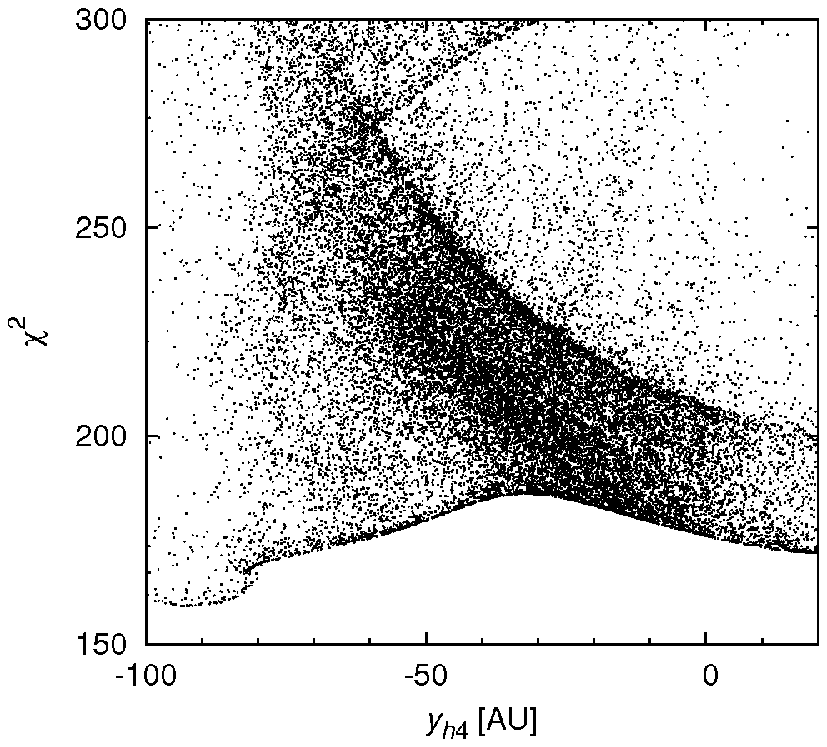}
\includegraphics[height=4.8cm]{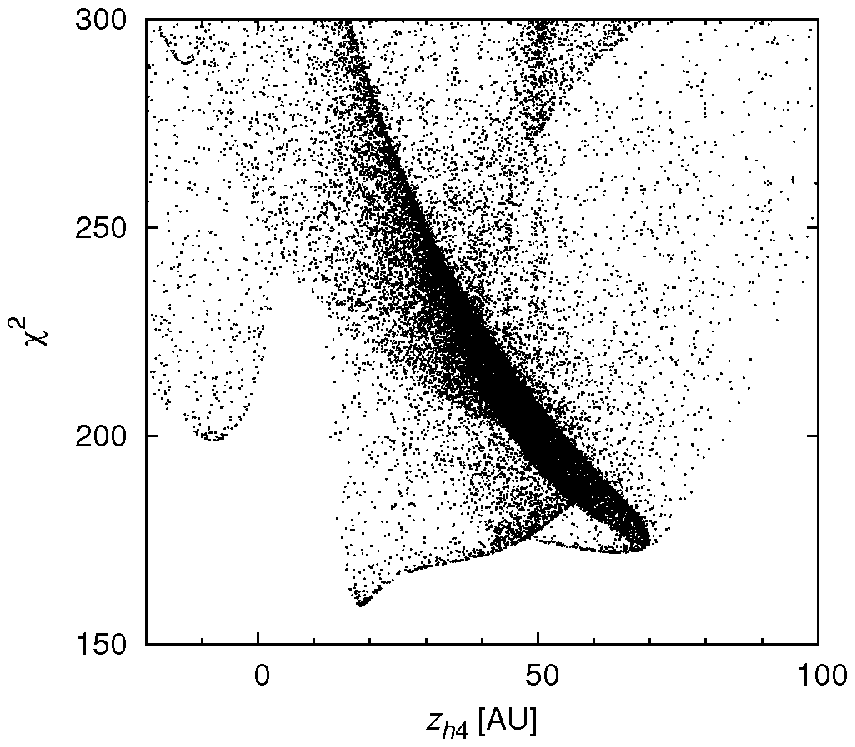}
\includegraphics[height=4.8cm]{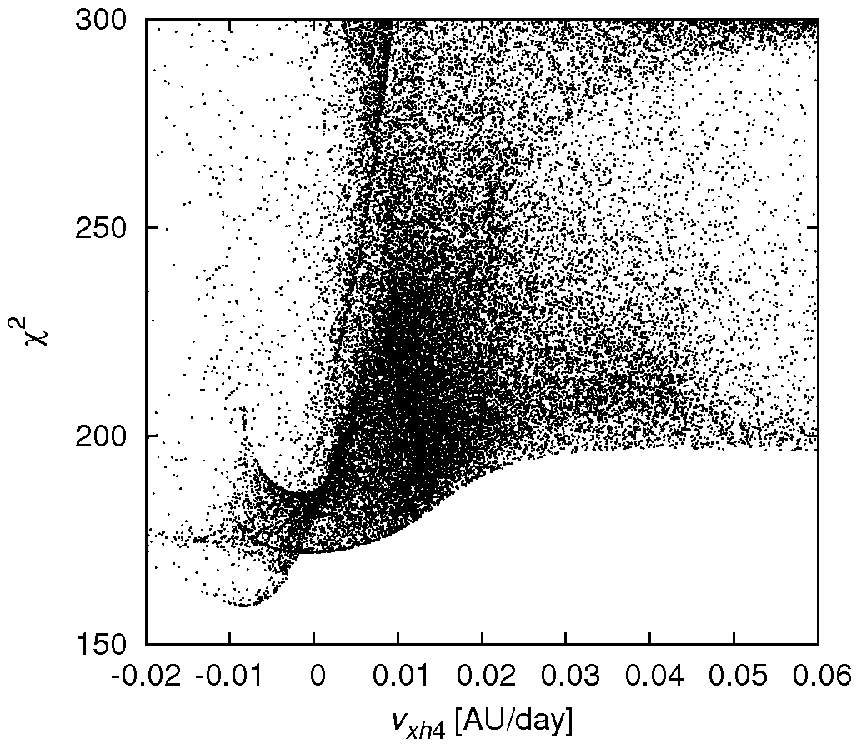}
\includegraphics[height=4.8cm]{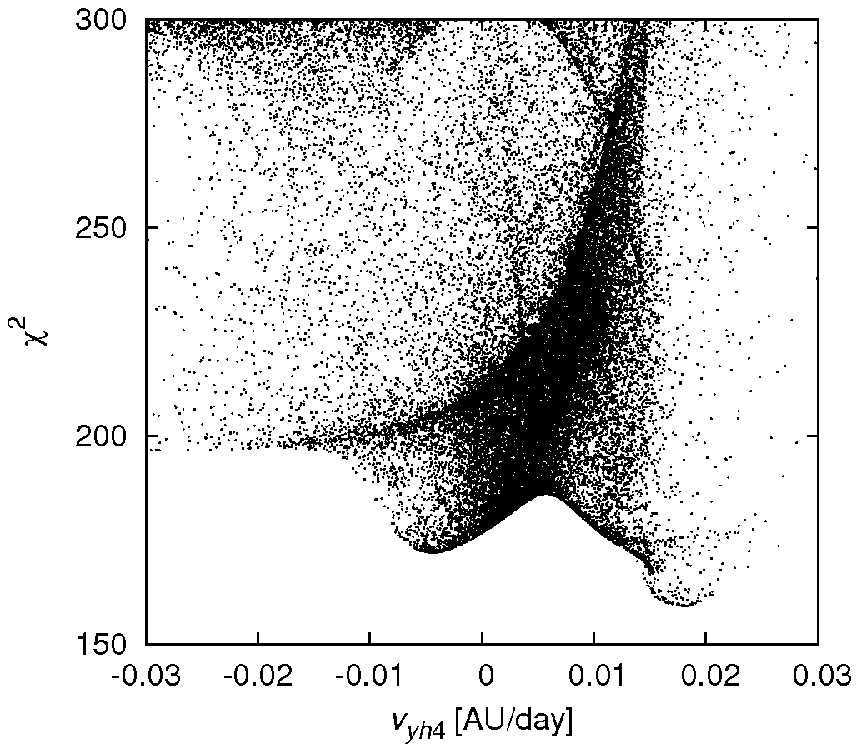}
\includegraphics[height=4.8cm]{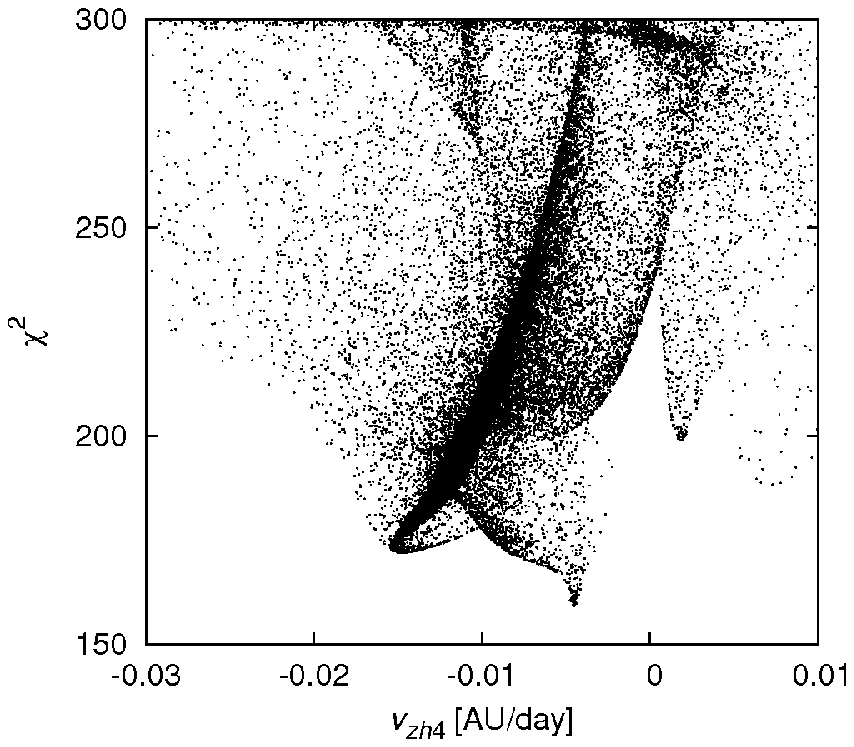}
\caption{Distribution of good solutions ($\chi^2 < 300$) in the space of free parameters:
$x_{h4}, y_{h4}, z_{h4}, v_{xh4},$ $ v_{yh4}, v_{zh4}$.}
\label{simplex_4TH_BODY9_simplex_x_chi2_DETAIL_eps}
\end{figure*}

As before, there are many solutions, which are statistically equivalent.
We present allowed solutions in Figure~\ref{simplex_4TH_BODY9_simplex_x_chi2_DETAIL_eps}
as plots $\chi^2$ versus a free parameter, with each dot representing one local minimum
found by simplex. Prominent concentrations of solutions in these plots
can be regarded as an indication of more probable solutions.
Only minority of trials were successful. Most of them were stopped too early
(at high~$\chi^2$) due to numerous local minima.

\begin{figure*}
\centering
\includegraphics[height=4.8cm]{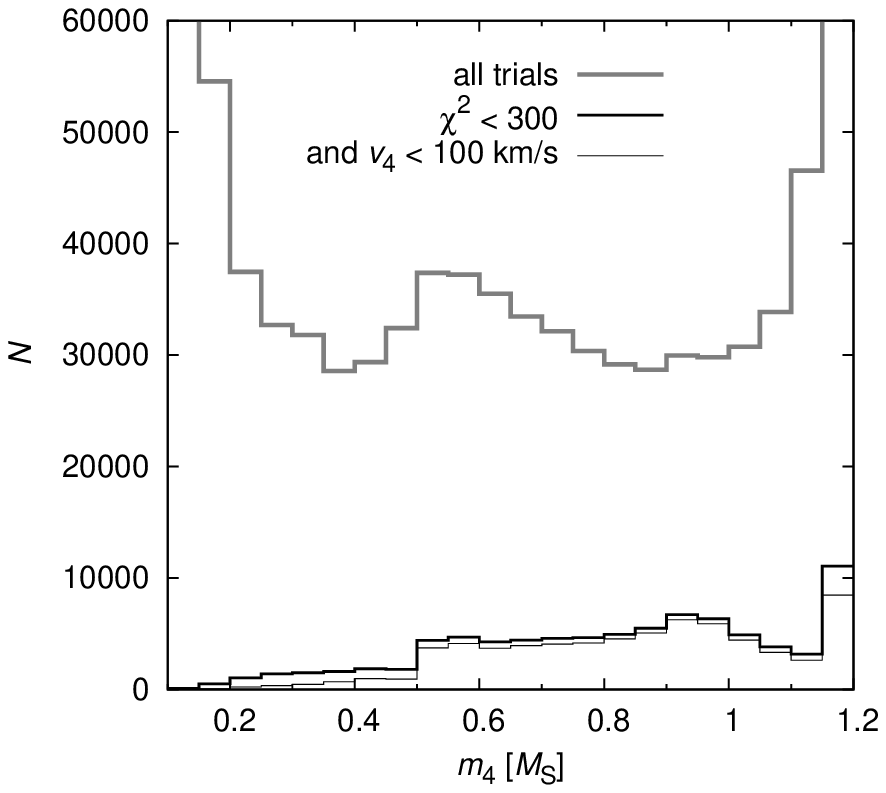}
\includegraphics[height=4.8cm]{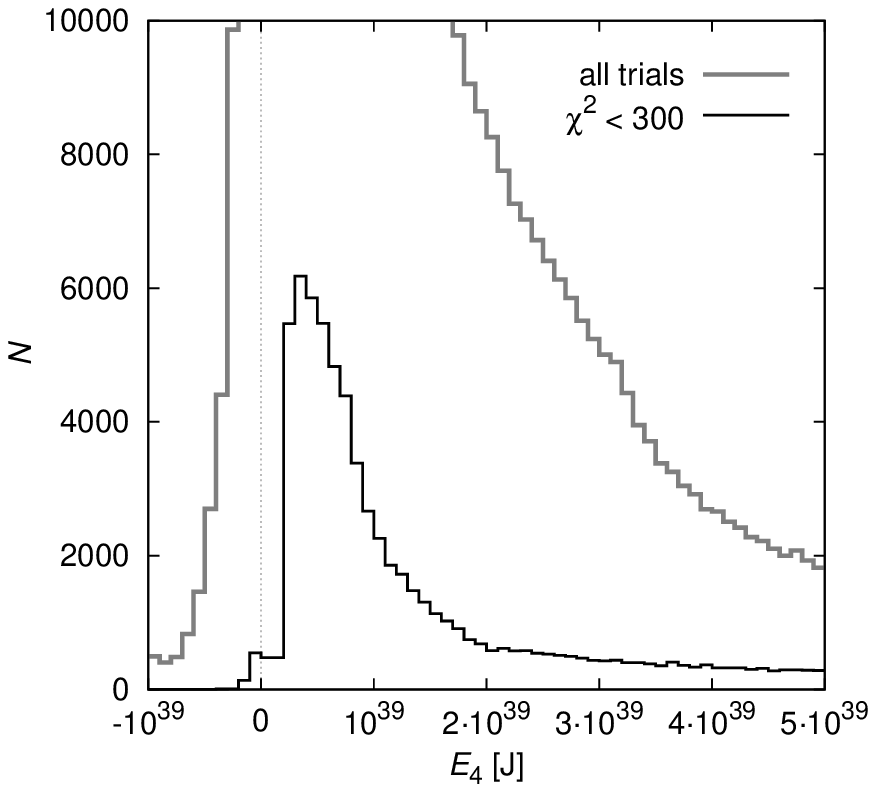}
\includegraphics[height=4.8cm]{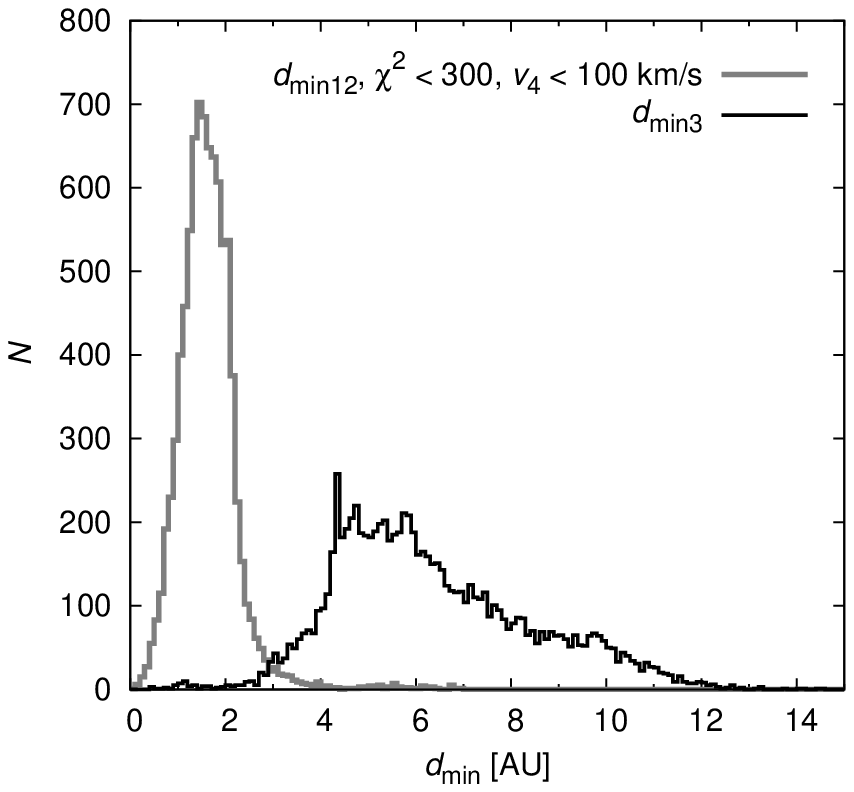}
\caption{Left: Histogram of masses~$m_4$ of all trials and good ones (with $\chi^2 < 300$
and lower absolute value of velocity $|v_4| < 100\,{\rm km}/{\rm s}$).
Middle: Histogram of total energies~$E_4$ of the 4th body at the epoch~$T_0$.
There is a strong preference for hyperbolic orbits ($E_4 > 0$), but one can
find approximately 1\,\% of orbits with negative energies.
Right: Histogram of minimum distances $d_{{\rm min}1+2}$, $d_{{\rm min}3}$ of the 4th body
from the (1+2) and 3rd body (only good trials are shown).
According to these distributions, it is more likely, that the 4th body
approached the (1+2) body closer than the 3rd body.
Median values are $\bar d_{{\rm min}1+2} = 1.6\,{\rm AU}$
and $\bar d_{{\rm min}3} = 5.9\,{\rm AU}$.}
\label{simplex_4TH_BODY9_m_hist_rozumne_vmax}
\end{figure*}

According to the histogram of masses~$m_4$ (Figure~\ref{simplex_4TH_BODY9_m_hist_rozumne_vmax}, left)
the values $m_4 < 0.5\,M_\odot$ are less probable and the histogram peaks around $m_4 = 0.9\,M_\odot$.
Note the simplex sometimes tends to `drift' to zero or large masses,
which leads to artificial peaks at the limits of the allowed interval.
The same applies to velocity~$v_4$.

Histogram of total energies~$E_4$ of the 4th body (Figure~\ref{simplex_4TH_BODY9_m_hist_rozumne_vmax}, middle)
shows a strong preference for hyperbolic orbits ($E_4 > 0$),
but elliptic orbits ($E_4 < 0$) also exist (with a 1\,\% probability
and slightly larger best $\chi^2 = 199$). The reason for this preference
stems from the fact that 3rd body orbit seems almost unperturbed
prior to 2000, so one needs rather a higher-velocity encounter of the 4th body
from larger initial distance.

Typical minimum distances between the 4th and 3rd body during an encounter are around
$d_{{\rm min}3} \simeq 6\,{\rm AU}$
and they are even smaller between the 4th body and (1+2) body
$d_{{\rm min}1+2} \simeq 1.5\,{\rm AU}$
(Figure~\ref{simplex_4TH_BODY9_m_hist_rozumne_vmax}, right).
They are of comparable size and consequently a simple impulse approximation,
i.e., an instantaneous change of orbital velocity,
cannot be used to link the two elliptic orbits of the 3rd body
(before and after the perturbation).
There are no good solutions (with $\chi^2 < 300$), which would lead to an escape
of the 3rd body.

\subsection{Observational limits of interferometry and CFHT imaging}\label{sec:observational_limits}

Postulating an existence of a 4th body, inferred from its gravitational
influence on the V505 Sgr system, we should check if this object could
have been directly observed in the past.

According to A.~Tokovinin (personal communication) the limit of recent
interferometric measurements can be approximated by a linear dependence
of brightness difference~$\delta y$ in Str\"omgren $y$ magnitudes
on angular separation~$d$ of the components:
$\Delta y = 4.7\,{\rm mag}$ at $d = 0.15\,{\rm arcsec}$
and
$\Delta y = 6.5\,{\rm mag}$ at $d = 1\,{\rm arcsec}$
Adaptive optics at CFHT can reach even fainter. The limit in $K$ band is given
by Rucinski et al. (2007), Fig.~7 as a non-linear dependence~$\Delta K(d)$.

\begin{figure*}
\centering
\includegraphics[height=5.2cm]{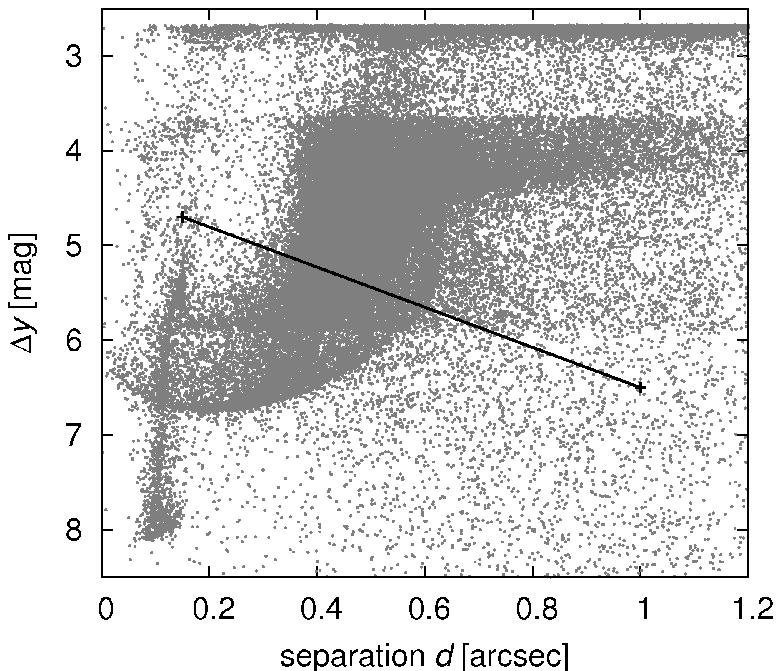}
\includegraphics[height=5.2cm]{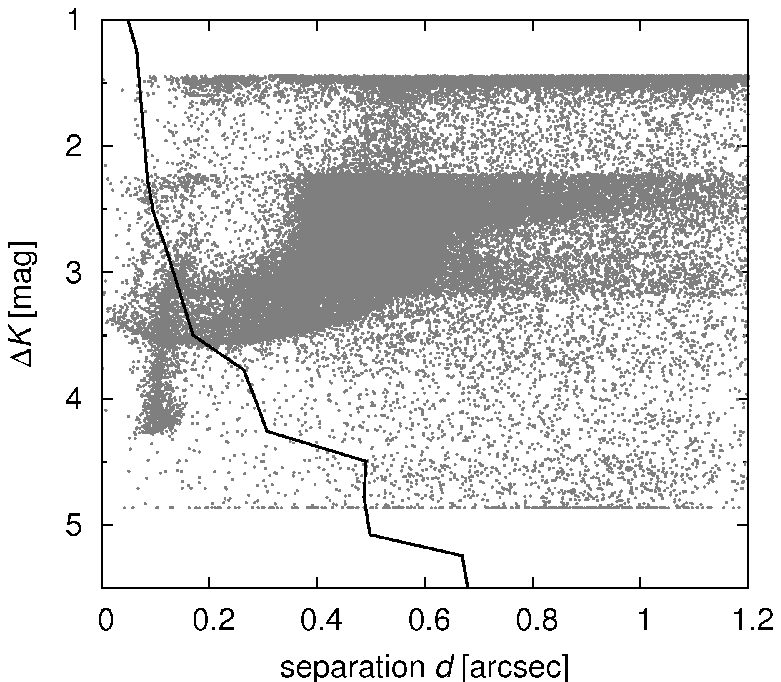}
\caption{Observational limits (black lines) of interferometric measurements (left panel)
and the CFHT imaging (right panel). Our solutions from Section~\ref{sec:simplex_4TH_BODY9}
are plotted as dots. Many of them are well below the observational limits.}
\label{obs_limit_CFHT}
\end{figure*}

We can easily select solutions from Section~\ref{sec:simplex_4TH_BODY9},
which fulfil {\em both\/} limits, albeit a lot of them is excluded by the CFHT limit
(see Figure~\ref{obs_limit_CFHT}).
Note we are not able to predict exact magnitudes or positions
of the 4th body, because there are still many solutions possible.

Note there is an object in the USNO-A2.0 catalogue, very close to V505 Sgr:
0750-19281506,
${\rm RA}_{\rm J2000} = 298.277034^\circ$,
${\rm DE}_{\rm J2000} = -14.603839^\circ$.
This corresponds to an angular separation $2.6$\,arcsec
and position angle~$235^\circ$
with respect to V505 Sgr, at the epoch of observation 1951.574.
The magnitudes 
$R = 10.9\,{\rm mag}$
and
$B = 11.8\,{\rm mag}$
are marked as uncertain
(since the object is located in the area flooded by light of V505 Sgr).
This is an interesting coincidence with "our" 4th body, but we doubt the source is real.
Moreover, if the brightness of the USNO source is correct within $\pm1\,{\rm mag}$,
it should be above the observational limits.

\subsection{Constraints from spectral lines radial-velocity measurements}

In previous Section~\ref{sec:optimalizace_4TH_BODY_7D_detail}, we tried to attribute
the observed high radial velocities to a hypothetic 4th component.
We thus have to ask a question: could the low-mass 4th component be visible in the spectrum?

To this end, we used a grid of synthetic spectra
based on Kurucz model atmospheres, which was calculated
and provided for general use by Dr.~J. Kub\'at
(for details of the calculations, cf., e.g., Harmanec et al. 1997).
We calculate synthetic spectra for 3 and 4 lights (stars)
and compare them with the spectrum observed by Tomkin (1992), Fig.~2.
This spectrum was taken at HJD = 2444862.588, close to the primary eclipse of the central binary,
which decreases the luminosity of the 1st component and thus weak narrow lines
of the 3rd (or 4th) component are more prominent.

Modelling of spectra (relative intensities) requires a number of parameters:
luminosities,
effective temperatures,
surface gravity,
rotational and radial velocities.
Luminosities of the known components (out of eclipse) are:
$L_1 = 26\,L_\odot$,
$L_2 =  3.8\,L_\odot$,
$L_3 =  2.1\,L_\odot$.
The amplitude of the lightcurve is $\Delta m = 1.1\,{\rm mag}$ (Chambliss et al. 1993).
The effective temperatures are approximately (Popper 1980):
$T_{{\rm eff}1} \simeq 9000\,{\rm K}$ (corresponding to A2~V spectral type),	
$T_{{\rm eff}2} \simeq 6000\,{\rm K}$ (F8~IV to G6--8~IV),			
$T_{{\rm eff}3} \simeq 6000\,{\rm K}$ (F8~V).
We assume the following values of the surface gravitational acceleration:
$\log g_1 = \log g_2 = 4.0$ (cgs units),
$\log g_3 = 4.5$ (valid for stars close to the main sequence).
Rotational velocities of the 1st and 2nd components, a semi-contact binary with
an orbital period 1.2~day, are synchronised by tidal lock and are of the order
$v_{{\rm rot}1} \simeq v_{{\rm rot}2} \simeq 100\,{\rm km}/{\rm s}$.
These are in concert with the observed width of broad spectral lines $\Delta\lambda = 6$\,\AA.
For the 3rd component, we assume a lower velocity $v_{{\rm rot}3} = 20\,{\rm km}/{\rm s}$,
usual for main-sequence stars. This matches the width of sharp lines.
Radial velocities of the 1st and 2nd components are close to zero
because of the eclipse proximity ($v_{{\rm rad}1} = v_{{\rm rad}2} \doteq 0$).

We assume the following reasonable parameters for the 4th component:
$T_4 = 4000\,{\rm K}$ or $5700\,{\rm K}$,
$\log g_4 = 4.0$,
$v_{{\rm rot}4} = 20\,{\rm km}/{\rm s}$.
We construct a $\chi^2$ metric
\begin{equation}
\chi^2 = \sum_{i=1}^{N_{\rm obs}} {(I_{\rm obs}[i] - I')^2\over\sigma_{\rm obs}[i]^2}\,,
\end{equation}
where $I_{\rm obs}[i]$ denote observed relative intensities,
$\sigma_{\rm obs}[i]$ associated uncertainties and
$I'$ is a sum of synthetic intensities weighted by luminosities
\begin{equation}
I' = {\sum_{j=1}^4 I'(T_{{\rm eff}j}, \log g_j, v_{{\rm rot}j}) \cdot L_j \over \sum_{j=1}^4 L_j}
\end{equation}
and of course Doppler shifted due to radial velocities
($\lambda' = \lambda_{\rm obs}[i] (1 - v_{{\rm rad}j}/c)$)
and interpolated to the required wavelengths $\lambda'$
using Hermite polynomials (Hill 1982).
We use a simple eclipse modelling: we decrease $L_1$ according
to the Pogson equation to get the observed total magnitude increase~$\Delta m$.
Errors $\sigma_{\rm obs}[i]$ were estimated from the scatter in small continua, $\sigma = 0.01$.
Artificially small errors $\sigma = 0.003$ were assigned to the measurements
in the cores of the narrow lines, in order to match precisely their depths.




We constructed a simplex algorithm (Press et al.~1997) with the following
free parameters:
$L_4$,
$v_{{\rm rad}3}$,
$v_{{\rm rad}4}$.
Other luminosities and radial velocities remain fixed.
This simplex is well-behaved and converges to final values almost
regardless of starting point. There is no reasonable improvement,
if we let all 8~parameters ($L_j$, $v_{{\rm rad}j}$) to be free.

\begin{figure}
\centering
\includegraphics[width=7.5cm]{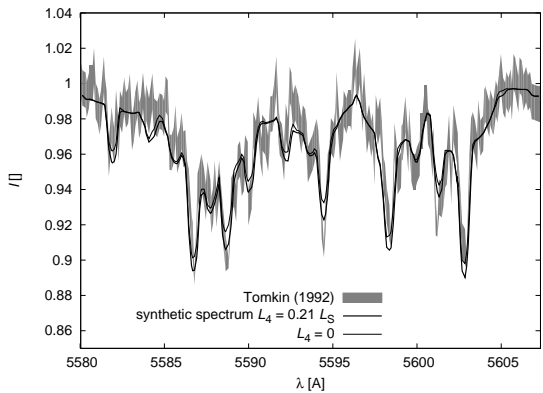}
\caption{A synthetic spectrum with intensity normalised to continuum for 4 and 3 lights (thick and thin lines)
and for the temperature of the 4th body equal to 4000\,K,
compared with the observed spectrum (gray line, which thickness corresponds to observational uncertainties),
taken from Tomkin (1992).
The wavelength range, which we fitted, was 5580\,\AA\ to 5607\,\AA.
The best solution for the luminosity of the 4th body is $L_4 = 0.22\,L_\odot$ (with $\chi^2 = 465$, $N = 239$)
The fit with 3~lights only is worse, with $\chi^2$ value equal to 542.
For $T_4 = 5700\,K$ we would have $L_4 = 0.43\,L_\odot$ and $\chi^2 = 495$.}
\label{3_and_4_lights}
\end{figure}

The results for two different temperatures are shown in Figure~\ref{3_and_4_lights}.
The best fit for $T_4 = 4000\,K$ is $L_4 = 0.22\,L_\odot$, and it is marginally
better that the fit with 3~lights only (i.e., with fixed $L_4 = 0$).
The luminosity corresponds roughly to the mass $m_4/M_\odot \propto (L_4/L_\odot)^{1/4} = 0.68$,
which seems reasonable with respect to the results in Section~\ref{sec:simplex_4TH_BODY9}.

Note we used $v_{{\rm rot}3} = 20\,{\rm km}/{\rm s}$ for rotational velocity of the 3rd body.
No reasonable solution was found for $v_{{\rm rot}3}$ as high as $100\,{\rm km}/{\rm s}$,
which would cause a strong rotational broadening and almost a `disappearance'
of spectral lines of the 3rd body. It means, that a low-mass 4th body {\em alone\/} cannot
produce deep sharp lines. We thus suspect, there is a blend of lines
in the spectrum of observed by Tomkin (1992), which may originate on the 3rd and 4th body,
with low and high radial velocities.
However, observations with high spectral resolution would be needed to resolve such blending.

\subsection{Constraints from the stellar evolution of the eclipsing binary}

\begin{figure*}
\centering
\includegraphics[width=9cm]{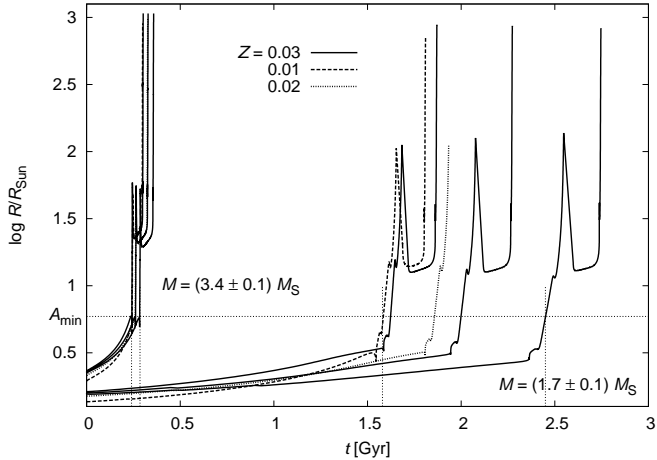}
\caption{Evolution of radii for stars with different masses ($M = 1.6, 1.7, 1.8$ and $3.3, 3.4, 3.5\,M_\odot$, $Z = 0.03$)
and metallicities ($M = 1.7$ and $3.4\,M_\odot$, $Z = 0.01, 0.02$).
This computation was performed by the program EZ (Paxton 2004).
The upper limit for the age of the 2nd component is then 1.5 to 2.5\,Gyr,
while the lower limit is between 0.24 and 0.29\,Gyr.}
\label{Rt_M1.7}
\end{figure*}

To assess the long-term evolution of V505 Sgr, we need some information
about the age of the system. 
An {\em upper\/} limit for the age can be estimated easily from masses of stars.
The semi-detached central binary (bodies 1 and 2) has a total mass $(3.4\pm0.1)\,M_\odot$.
In order to evolve into the current stage, when the 2nd {\em lighter\/} component fills
its Roche lobe, the original mass of the 2nd star had to be at least slightly
larger than half of the total mass, i.e., $M_2 > 1.7\,M_\odot$. The evolution
of radius is shown in Figure~\ref{Rt_M1.7}; we are mainly concerned with
the large increase of radius, when the star leaves main sequence.
Given the uncertainties of the masses and unknown metallicities,
the upper limit for the age is $(2.0\pm0.5)\,{\rm Gyr}$.

In order to find a lower limit, we have to check a minimum separation
of the components first (cf.~Eq.~\ref{eq:A_M_1} and Figure~\ref{vzdalenost_V505Sgr}).
A minimum separation occurs when $M_1 = 0.5 K$, in our case $A_{\rm min} = (5.9\pm0.1)\,R_\odot$.
This value is larger than the radius of a $3.4\,M_\odot$ star
during the whole evolution on the main sequence. Thus the mass transfer
had to start later, in the red-giant phase.

The maximum mass of the 2nd star had to be slightly below the total mass, i.e., $M_2 < 3.4\,M_\odot$.
According to the $R(t)$ dependence (Figure~\ref{Rt_M1.7}), the red-giant
phase starts at the age of $(0.26\pm0.03)\,{\rm Gyr}$, which could be considered
as a {\em lower\/} limit for the age of the V505 Sgr system.

\section{Conclusions}\label{sec:conclusions}

Generally speaking, we are able to explain the observed orbit of the 3rd body
together minima timings and radial velocities by a low-mass 4th body, which encounters the observed
three bodies with a suitable geometry. There is no unique solution, but rather
a set of allowed solutions for the trajectory of the hypothetic 4th body.
It is quite difficult to find a solution for both speckle-interferometry and
light-time effect data. There are a few systematic discrepancies at the $2\sigma$~level,
which cause the likelihood of the hypothesis to be low. Possibly, realistic uncertainties $\sigma_{\rm sky}$
are slightly larger (by a factor 1.5) than the errors estimated by us.

Of course, there are other hypotheses, which do not need a 4th body at all
(a sudden mass transfer, Applegate's mechanism, etc.),
but none of them provides a unified solution for {\em all\/} observational data
we have for V505~Sgr.

Further observations of the light time-effect during the next decade
can significantly constrain the model.
A new determination of the systemic velocity of V505 Sgr
may confirm, that the change in the $O-C$ data after 2000
resulted from an external perturbation. (Tomkin's (1992)
value was $(1.9\pm 1.4)\,{\rm km}/{\rm s}$.)
Spectroscopic measurements of the indicative sharp lines would be also
very helpful to resolve the problem with radial velocities mentioned in the text.


If we indeed observe the V505 Sgr system by chance during the phase
of a close encounter with a 4th star, we can imagine several scenarios
for its origin:

\begin{enumerate}
\item A random passing star approaching V505 Sgr on a hyperbolic orbit.
The problem of this scenario is a very low number density of stars.
If we take the value $n_\star \simeq 0.073\,{\rm pc}^{-3}$ from the solar vicinity
(Fern\'andez 2005), the mean velocity with respect to other stars
of the order $v_{\rm rel} \simeq 10\,{\rm km}/{\rm s}$
and the required minimum distance of the order $d \simeq 10^2\,{\rm AU}$,
we end up with a mean time between two encounters
$\tau \simeq {1 / (n_\star v_{\rm rel} d^2)} \simeq 10^{12}\,{\rm yr}$,
thus an extremely unlikely event.


\item A loosely bound star on a highly eccentric orbit, with the same age
as other three components of V505 Sgr.
Unfortunately, there is a large number of revolutions and encounters ($10^2$ to $10^5$)
over the estimated age of V505 Sgr and the system practically cannot remain stable over this time scale
(Valtonen \& Mikkola 1991).


\item A more tightly bound star on a lower-eccentricity orbit,
which experienced some sort of a late instability, induced by long-term
evolution due to galactic tides, distant passing stars, which shifted an initially
stable configuration into an unstable state, e.g., driven by mutual gravitational
resonances between components.
The problem in this case is that tightly bound orbits of the 4th body
are very rare in our simulations, thus seem improbable.



\end{enumerate}

None of the scenarios is satisfactory.
Nevertheless, we find the 4th-body hypothesis the only one
which is able to explain all available observations.
Clearly, more observations and theoretical effort is needed
to better understand the V505 Sagittarii system.


\begin{acknowledgements}
This research has made use of the WDS Catalog maintained at the U.S. Naval Observatory
and Aladin and Simbad services.
We thank W.I.~Hartkopf for sending us recent positional data of double stars,
S.~Rucinski and E.~Malogolovets for providing us with their measurements
of the visual component of V505~Sgr (by CFHT and SAO BTA telescopes)
and L.~\v Smelcer for his photometry of V505~Sgr.
We acknowledge the use of a grid of synthetic spectra prepared by J.~Kub\'at
and we thank P.~Harmanec for some advice on the construction of synthetic spectra
for multiple systems.
This work was supported by the Czech Science Foundation (grant no.~P209/10/0715)
and the Research Programme MSM0021620860 of the Czech Ministry of Education.
TP~acknowledges support from the EU in the FP6 MC ToK project MTKD-CT-2006-042514.

\end{acknowledgements}


\end{document}